\documentclass{SciPost}

\hypersetup{
 colorlinks,
 linkcolor={red!50!black},
 citecolor={blue!50!black},
 urlcolor={blue!80!black},
 pdftitle={From quantum fluctuations to galaxy power spectrum multipoles},
 pdfauthor={Nhat-Minh Nguyen},
 pdfsubject={Linear structure formation, galaxy bias, redshift-space multipoles,
 one-loop EFT, and FFTLog},
 pdfkeywords={large-scale structure, galaxy bias, redshift-space distortions,
 multipoles, Gaussian covariance, effective field theory, FFTLog}
}

\usepackage[bitstream-charter]{mathdesign}
\DeclareSymbolFont{usualmathcal}{OMS}{cmsy}{m}{n}
\DeclareSymbolFontAlphabet{\mathcal}{usualmathcal}

\DeclareMathVersion{cmfig}
\SetSymbolFont{operators}{cmfig}{OT1}{cmr}{m}{n}
\SetSymbolFont{letters}{cmfig}{OML}{cmm}{m}{it}
\SetSymbolFont{symbols}{cmfig}{OMS}{cmsy}{m}{n}
\SetSymbolFont{largesymbols}{cmfig}{OMX}{cmex}{m}{n}
\newenvironment{cmfigurefont}
 {\fontfamily{cmr}\selectfont\mathversion{cmfig}}
 {}

\usepackage{microtype}
\usepackage{mathtools,bm}
\usepackage{booktabs,tabularx,array,multirow}
\usepackage{enumitem}
\usepackage{needspace}
\usepackage{placeins}
\usepackage{tikz}
\usetikzlibrary{arrows.meta,positioning,calc,shapes.geometric}
\usepackage[most]{tcolorbox}
\usepackage{xurl}
\usepackage{orcidlink}
\usepackage[nameinlink,capitalise]{cleveref}
\graphicspath{{figs/}}
\fancypagestyle{SPstyle}{
\fancyhf{}
\lhead{\colorbox{scipostblue}{\bf \color{white} ~SciPost Physics Lecture Notes }}
\rhead{{\bf \color{scipostdeepblue} ~Submission }}

\fancyfoot[C]{\textbf{\thepage}}
}

\colorlet{Ink}{scipostdeepblue}
\colorlet{LinkBlue}{scipostblue}
\colorlet{DeepGreen}{scipostdeepblue}
\colorlet{Plum}{scipostdeepblue}
\colorlet{SoftBlue}{scipostblue!6}
\colorlet{SoftGreen}{scipostdeepblue!7}
\colorlet{SoftGray}{black!4}

\definecolor{BoxForest}{HTML}{2F5D50}
\definecolor{BoxSeaBlue}{HTML}{2A6F89}
\definecolor{BoxAmber}{HTML}{91651B}
\definecolor{BoxSand}{HTML}{F5F0E6}

\titleformat{\subsection}{\large\bfseries\color{Ink}}{\thesubsection}{0.7em}{}
\titleformat{\subsubsection}{\normalsize\bfseries\color{Ink}}{\thesubsubsection}{0.65em}{}
\titlespacing*{\subsection}{0pt}{2.2ex plus .5ex minus .2ex}{0.8ex}
\titlespacing*{\subsubsection}{0pt}{1.7ex plus .4ex minus .2ex}{0.5ex}
\setlist{itemsep=2pt,topsep=4pt,leftmargin=*}
\numberwithin{equation}{section}

\newtcolorbox{objectives}{
 breakable,enhanced,title={Objectives},fonttitle=\bfseries,
 coltitle=white,colbacktitle=BoxSeaBlue,colback=BoxSeaBlue!6,colframe=BoxSeaBlue,
 boxrule=0.6pt,arc=0.5mm,left=2mm,right=2mm,top=1.2mm,bottom=1.2mm}
\newtcolorbox{keybox}[1]{
 breakable,enhanced,title={Summary: #1},fonttitle=\bfseries,
 coltitle=white,colbacktitle=BoxForest,colback=BoxForest!6,colframe=BoxForest,
 boxrule=1.1pt,arc=0.5mm,left=2mm,right=2mm,top=1.2mm,bottom=1.2mm}
\newtcolorbox{warningbox}[1]{
 breakable,enhanced,title={Remark: #1},fonttitle=\bfseries,
 coltitle=white,colbacktitle=BoxAmber,colback=BoxAmber!8,colframe=BoxAmber,
 boxrule=0.8pt,arc=0.5mm,left=2mm,right=2mm,top=1.2mm,bottom=1.2mm}
\newtcolorbox{readingbox}[1]{
 breakable,enhanced,title={Supplementary: #1},fonttitle=\bfseries,
 coltitle=BoxAmber!75!black,colbacktitle=BoxAmber!12,
 colback=BoxSand,colframe=BoxAmber!35,
 boxrule=0.55pt,arc=0.5mm,left=2mm,right=2mm,top=1.2mm,bottom=1.2mm,
 fontupper=\small}
\tcbset{seablueworkbox/.style={
 breakable,enhanced,fonttitle=\bfseries,
 coltitle=BoxSeaBlue!75!black,colbacktitle=BoxSeaBlue!12,
 colback=white,colframe=BoxSeaBlue!45,
 boxrule=0.5pt,arc=0.5mm,left=2mm,right=2mm,top=1.2mm,bottom=1.2mm}}
\newtcolorbox{derivationbox}[1]{seablueworkbox,title={Derivation: #1}}

\newcommand{\dd}{\mathrm{d}}
\newcommand{\ii}{\mathrm{i}}
\newcommand{\vx}{\bm{x}}
\newcommand{\va}{\bm{a}}
\newcommand{\vs}{\bm{s}}
\newcommand{\vq}{\bm{q}}
\newcommand{\vk}{\bm{k}}
\newcommand{\vp}{\bm{p}}
\newcommand{\vu}{\bm{u}}
\newcommand{\vPsi}{\bm{\Psi}}
\newcommand{\nhat}{\hat{\bm{n}}}
\newcommand{\khat}{\hat{\bm{k}}}
\newcommand{\Plin}{P_{\mathrm L}}
\newcommand{\dirac}{\delta_{\mathrm D}}
\newcommand{\cH}{\mathcal H}
\newcommand{\Gtwo}{\mathcal G_2}
\newcommand{\Gthree}{\Gamma_3}
\newcommand{\Cov}{\operatorname{Cov}}
\newcommand{\Lpoly}{\mathcal L}

\newcommand{\intk}{\int\!\frac{\dd^3k}{(2\pi)^3}}
\newcommand{\intq}{\int\!\frac{\dd^3q}{(2\pi)^3}}
\newcommand{\Order}{\mathcal O}
\newcommand{\code}[1]{\texttt{#1}}
\newcommand{\lectureref}[1]{\hyperref[lec:#1]{Lecture~#1}}
\newcommand{\extensionref}{\hyperref[ext:one-loop]{advanced extension}}

\newcommand{\preludetitle}[1]{%
 \clearpage
 \phantomsection
 \addcontentsline{toc}{section}{Prelude: #1}
 \markboth{Prelude: #1}{Prelude: #1}
 \setcounter{section}{0}
 \setcounter{subsection}{0}
 \setcounter{subsubsection}{0}
 \setcounter{equation}{0}
 \begin{center}
 {\LARGE\bfseries\color{Ink} Prelude: #1}\par
 \vspace{5pt}\color{Ink!45}\rule{0.92\textwidth}{0.5pt}
 \end{center}
 \vspace{0.5em}
}

\newcommand{\lecturetitle}[2]{%
 \clearpage
 \phantomsection
 \label{lec:#1}
 \addcontentsline{toc}{section}{Lecture #1: #2}
 \markboth{Lecture #1: #2}{Lecture #1: #2}
 \setcounter{section}{#1}
 \setcounter{subsection}{0}
 \setcounter{subsubsection}{0}
 \setcounter{equation}{0}
 \begin{center}
 {\LARGE\bfseries\color{Ink} Lecture #1: #2}\par
 \vspace{5pt}\color{Ink!45}\rule{0.92\textwidth}{0.5pt}
 \end{center}
 \vspace{0.5em}
}

\newcommand{\extensiontitle}[1]{%
 \clearpage
 \phantomsection
 \label{ext:one-loop}
 \addcontentsline{toc}{section}{#1}
 \markboth{#1}{#1}
 \setcounter{section}{4}
 \setcounter{subsection}{0}
 \setcounter{subsubsection}{0}
 \setcounter{equation}{0}
 \begin{center}
 {\small\bfseries\scshape\color{LinkBlue} Beyond the VSOA10 lectures}\par\vspace{3pt}
 {\LARGE\bfseries\color{Ink} #1}\par
 \vspace{5pt}\color{Ink!45}\rule{0.92\textwidth}{0.5pt}
 \end{center}
 \vspace{0.5em}
}

\begin{document}
\pagestyle{SPstyle}

\begin{center}{\Large \textbf{\color{scipostdeepblue}{
From quantum fluctuations to galaxy power spectrum multipoles\\
}}}\end{center}

\begin{center}\textbf{
Nhat-Minh Nguyen\orcidlink{0000-0002-2542-7233}\textsuperscript{1,2$\star$}
}\end{center}

\begin{center}
{\bf 1} Center for Data-Driven Discovery, Kavli IPMU (WPI), UTIAS,
The University of Tokyo, Kashiwa, Chiba 277-8583, Japan
\\
{\bf 2} Kavli IPMU (WPI), UTIAS, The University of Tokyo,
5-1-5 Kashiwanoha, Kashiwa, Chiba 277-8583, Japan
\\[\baselineskip]
$\star$ \href{mailto:nhat.minh.nguyen@ipmu.jp}{\small nhat.minh.nguyen@ipmu.jp}
\end{center}

\section*{\color{scipostdeepblue}{Abstract}}
\textbf{\boldmath{%
These notes trace large-scale structure from primordial curvature perturbations generated by inflationary quantum fluctuations to galaxy power-spectrum multipoles.
Three core lectures develop the linear matter power spectrum, spherical and anisotropic collapse, galaxy bias, redshift-space distortions, the Kaiser model, and multipole estimators with Gaussian covariance.
The extension develops nonlinear bias and the one-loop effective field theory model used in full-shape analyses.
Derivations are explicit;
appendices collect longer calculations and solutions.
The core lectures assume undergraduate-level cosmology;
the extension assumes familiarity with perturbation theory.
}}

\vspace{\baselineskip}

\noindent\textcolor{white!90!black}{%
\fbox{\parbox{0.975\linewidth}{%
\textcolor{white!40!black}{\begin{tabular}{lr}%
 \begin{minipage}{0.6\textwidth}%
 {\small Copyright attribution to authors. \newline
 This work is a submission to SciPost Physics Lecture Notes. \newline
 License information to appear upon publication. \newline
 Publication information to appear upon publication.}
 \end{minipage} & \begin{minipage}{0.4\textwidth}
 {\small Received Date \newline Accepted Date \newline Published Date}%
 \end{minipage}
\end{tabular}}
}}
}

\vspace{\baselineskip}
\tableofcontents\thispagestyle{fancy}
\vspace{\baselineskip}

\section*{\color{scipostdeepblue}{Structure of these notes}}
\addcontentsline{toc}{section}{Structure of these notes}

The prelude introduces density fluctuations as a Gaussian random field and defines their power spectrum.
Lectures 1--3 are the taught VSOA10 sequence.
They develop a complete leading-order (tree-level) model for the galaxy power-spectrum multipoles, their Gaussian covariance in a periodic box, and the generalization of this covariance to a finite survey window, following Ref.~\cite{WadekarScoccimarro}.
The advanced extension develops the next-to-leading (one-loop) model for biased tracers in perturbation theory and the effective field theory of large-scale structure.
Appendix~\ref{app:master-integrals} derives its FFTLog evaluation, as implemented in \code{CLASS-PT}~\cite{ClassPT}, and states the minimal numerical checks.

\begin{center}
\begin{cmfigurefont}
\begin{tikzpicture}[node distance=5mm and 5mm,>=Latex,
 every node/.style={font=\small}]
\tikzset{stage/.style={draw=LinkBlue!70,rounded corners=2pt,fill=SoftBlue,
 minimum height=10mm,minimum width=27mm,align=center},
 stageii/.style={stage,draw=Ink,line width=0.8pt,fill=SoftGreen},
 stageiii/.style={stage,draw=Ink!55,line width=0.8pt,fill=SoftGray}}
\node[stage] (prim) {primordial\\spectrum};
\node[stage,right=of prim] (lin) {linear matter\\field};
\node[stage,right=of lin] (web) {collapse and\\cosmic web};
\node[stage,right=of web] (gal) {galaxy field\\in redshift space};
\node[stage,below=of gal] (mult) {multipoles and\\covariance};
\node[stageii,left=of mult] (eft) {advanced extension\\one-loop EFT};
\node[stageiii,left=of eft] (fft) {Appendix C\\FFTLog evaluation};
\draw[->,thick,LinkBlue] (prim)--(lin);
\draw[->,thick,LinkBlue] (lin)--(web);
\draw[->,thick,LinkBlue] (web)--(gal);
\draw[->,thick,LinkBlue] (gal)--(mult);
\draw[->,thick,Ink] (mult)--(eft);
\draw[->,thick,Ink] (eft)--(fft);
\end{tikzpicture}
\end{cmfigurefont}
\end{center}

\noindent Each lecture opens with a sea-blue \emph{Objectives} box.
Forest-green boxes carry \emph{Summary:} headings for results worth retaining.
Pale-sand boxes carry \emph{Supplementary:} material that may be skipped on a first reading.
Lightly ruled sea-blue boxes carry \emph{Derivation:} headings for calculations worked through in full.
Amber boxes carry \emph{Remark:} headings that state an assumption, limitation, or convention needed to interpret the adjacent result.

\noindent Reviews of the main topics are Refs.~\cite{Bernardeau} (nonlinear LSS and perturbation theory), \cite{BiasReview} (galaxy bias), \cite{Hamilton1998} (linear redshift-space distortions), and \cite{IvanovEFT} (EFTofLSS, renormalization and infrared resummation).
Three recent lecture notes are complementary:
Ref.~\cite{Uhlemann} emphasizes dark-matter dynamics, statistics and halo phenomenology, Ref.~\cite{Hahn} connects Gaussian fields and Lagrangian perturbation theory to simulation initial conditions and integrators, and Ref.~\cite{IvanovGGI} develops the one-loop galaxy model from effective field theory principles.
These notes proceed from primordial curvature to the multipoles measured by a spectroscopic survey.
The emphasis is on the transfer function, the estimator and its Gaussian covariance and mode count, and the FFTLog evaluation of biased redshift-space loops.
Each numbered lecture ends with a few exercises whose compact solutions are provided in \cref{app:answers}.

\preludetitle{the density field and its power spectrum}

\begin{objectives}
Students should be able to
\begin{itemize}
\item define density fluctuations as a scalar field;
\item move between the position and Fourier representations of that field;
\item describe density fluctuations as a Gaussian random field;
\item write down the power spectrum of a statistically homogeneous and isotropic Gaussian random field.
\end{itemize}
\end{objectives}

\subsection{The matter density contrast}

\begin{figure}[t]
\centering
\includegraphics[width=0.96\textwidth]{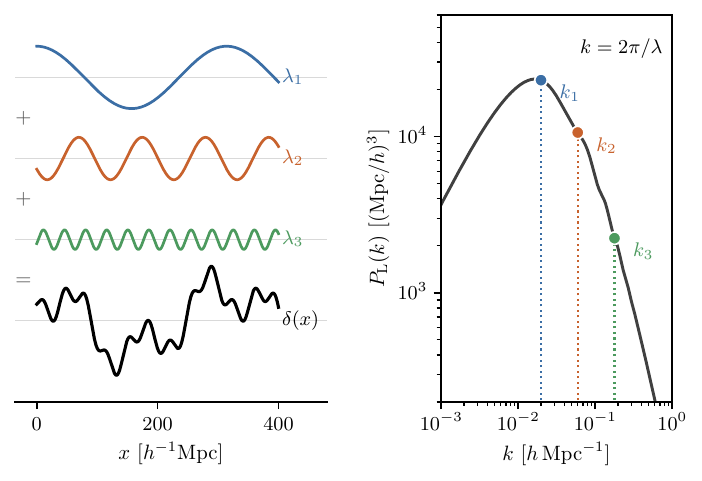}
\caption{Fourier mode decomposition.
\emph{Left:} three sinusoids of decreasing wavelength summed to form a realization $\delta(x)$.
Equation~\eqref{eq:fourier} gives this decomposition for a cosmological scalar field;
its coefficients are the Fourier-mode amplitudes.
\emph{Right:} the locations of the three wavenumbers on the linear matter power spectrum, with $k=2\pi/\lambda$.
A long wavelength $\lambda$ corresponds to a low-$k$ mode and carries the most power in this example, accounting for the dominant first mode in the left panel.}
\label{fig:fourier}
\end{figure}

We assume a spatially flat expanding universe;
flatness is an observational input rather than a result derived here.
Positions are labelled by comoving coordinates $\vx$, so physical separations are $a(t)\vx$ and a particle following the Hubble flow has fixed $\vx$.
We write the matter density as
\begin{equation*}
 \rho_m(\vx,t)=\bar\rho_m(t)\,[1+\delta_m(\vx,t)],
\end{equation*}
and define its fluctuation about the homogeneous background by
\begin{equation}
 \delta_m\equiv\frac{\rho_m-\bar\rho_m}{\bar\rho_m}.
 \label{eq:defdelta}
\end{equation}
The density contrast is dimensionless and has zero spatial mean by construction.
Overdense regions have $\delta>0$, whereas underdense regions have $\delta<0$.
The physical bound $\rho_m\geq0$ implies $\delta\geq-1$, and the linear regime corresponds to $|\delta|\ll1$.
We define the Fourier transform pair by
\begin{equation}
 \delta(\vk)=\int\dd^3x\,e^{-\ii\vk\cdot\vx}\delta(\vx),
 \qquad
 \delta(\vx)=\intk e^{\ii\vk\cdot\vx}\delta(\vk),
 \label{eq:fourier}
\end{equation}
in which a mode of comoving wavenumber $k$ has comoving wavelength $\lambda=2\pi/k$;
hence low $k$ corresponds to a large comoving scale.
Reality is a constraint on the Fourier coefficients rather than an extra assumption.
Conjugating the forward transform gives
\[
 \delta^*(\vk)
 =\left[\int\dd^3x\,e^{-\ii\vk\cdot\vx}\delta(\vx)\right]^*
 =\int\dd^3x\,e^{+\ii\vk\cdot\vx}\delta(\vx)
 =\delta(-\vk).
\]
The last equality follows by replacing $\vk$ with $-\vk$ in the forward transform.
Thus
\begin{equation}
 \delta(-\vk)=\delta^*(\vk).
 \label{eq:hermitian}
\end{equation}
Equivalently, changing the integration variable $\vk\to-\vk$ in the conjugate inverse transform returns the original real field.
This Hermitian relation is why a full Fourier shell contains twice as many entries as independent complex modes.
This distinction enters the covariance derivation of \lectureref{3}.
\cref{fig:fourier} illustrates the Fourier decomposition.

\subsection{Gaussian random fields and the power spectrum}

Statistical homogeneity and isotropy define the power spectrum of any random field.
Homogeneity enforces wavevector conservation, whereas isotropy restricts the remaining dependence to the magnitude $k$:
\begin{equation}
 \big\langle\delta(\vk)\delta(\vk')\big\rangle
 =(2\pi)^3\dirac(\vk+\vk')P(k).
 \label{eq:powerdef}
\end{equation}
The two-point definition applies to Gaussian and non-Gaussian fields independently of the dynamics that produced them.
In \lectureref{3} it will describe a biased galaxy field;
in the \extensionref{} it will describe a nonlinear matter field.

For a Gaussian field, the two-point function contains all statistical information, and Wick's theorem determines every higher correlator from it.
This Gaussian closure underlies the covariance in \lectureref{3} and the one-loop power spectrum in the \extensionref{}.
The power spectrum therefore measures the variance of Fourier modes, and it carries units of volume under the convention of \cref{eq:fourier}.
We also use the dimensionless power per logarithmic interval, $\Delta^2(k)\equiv k^3P(k)/(2\pi^2)$.
The factor $k^3$ counts the three-dimensional modes in a logarithmic shell.
Consequently, a small value of $P(k)$ at high $k$ can still contribute substantially to the variance.

\begin{derivationbox}{homogeneity and isotropy}
We translate both positions by a constant $\va$, and statistical homogeneity leaves the correlator unchanged while each Fourier mode gains a phase.
Therefore
\[
 \langle\delta(\vk)\delta(\vk')\rangle
 =e^{\ii(\vk+\vk')\cdot\va}
 \langle\delta(\vk)\delta(\vk')\rangle
\]
must hold for every $\va$.
The phase is independent of $\va$ only when $\vk+\vk'=0$, which is enforced by $\dirac(\vk+\vk')$.
Substituting \cref{eq:powerdef} into the inverse transforms then removes one momentum integral:
\begin{align}
 \langle\delta^2(\vx)\rangle
 &=\intk\int\!\frac{\dd^3k'}{(2\pi)^3}
 e^{\ii(\vk+\vk')\cdot\vx}
 (2\pi)^3\dirac(\vk+\vk')P(k)\nonumber\\
 &=\int\frac{\dd^3k}{(2\pi)^3}P(k)
 =\int_0^\infty\frac{\dd k\,k^2}{2\pi^2}P(k)
 =\int\dd\ln k\,\Delta^2(k).
 \label{eq:deltadimless}
\end{align}
No smoothing window has been applied.
The middle equality uses isotropy to perform the solid angle integral;
the last identifies the variance contributed per logarithmic interval.
\end{derivationbox}

\lecturetitle{1}{From primordial fluctuations to the linear matter power spectrum}

\begin{objectives}
By the end of this lecture a student should be able to
\begin{itemize}
\item distinguish sub-horizon from super-horizon modes and describe their evolution;
\item derive the linear growth equation from the continuity, Euler and Poisson equations;
\item distinguish $D_+$, normalized at early times, from $\widehat D_+$, normalized to unity today;
\item read the shape of $\Plin(k)$ as the combined record of primordial initial conditions and subsequent transfer physics.
\end{itemize}
\end{objectives}

\subsection{Modes and the horizon}

\begin{figure}[t]
\centering
\includegraphics[width=0.92\textwidth]{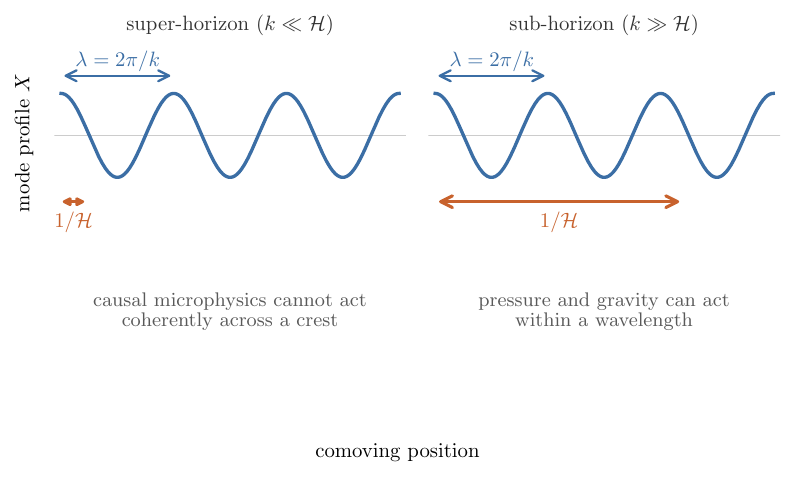}
\caption{Comoving wavelength relative to the Hubble radius.
\emph{Left:} an illustrative profile $X(x)$ when $k\ll\cH$;
the horizontal bar shows that the comoving Hubble radius $1/\cH$ is shorter than one wavelength, so causal microphysics cannot couple the full mode.
\emph{Right:} the same fixed comoving mode after $1/\cH$ exceeds several wavelengths, allowing gravity to act coherently across each wavelength.
The wavevector $\vk$ is unchanged because it is a comoving mode label;
only $1/\cH$ changes between the panels.}
\label{fig:wavelength}
\end{figure}

We define conformal time $\tau$ so that light travels one unit of comoving distance in one unit of $\tau$.
This makes the comoving Hubble radius $1/\cH$, with $\cH\equiv aH$, the scale separating what is in causal contact from what is not.
A mode is \emph{sub-horizon} when $k\gg\cH$, so that its comoving wavelength $\lambda=2\pi/k$ fits many times within $1/\cH$, and \emph{super-horizon} when $k\ll\cH$.
\cref{fig:wavelength} illustrates this horizon distinction.
Microphysical processes cannot act coherently across a super-horizon wavelength.
For the adiabatic growing mode, the comoving curvature perturbation $\mathcal R$ introduced in \cref{sec:primordial-curvature} is conserved.
The density contrast is gauge-dependent on these scales:
changing the time slicing changes the hypersurface on which densities are compared and therefore changes $\delta$.
It should not generically be described as frozen.
Once a mode is well inside the horizon, pressure and gravity can act across one wavelength and the matter perturbation evolves dynamically.

\begin{figure}[t]
\centering
\includegraphics[width=0.92\textwidth]{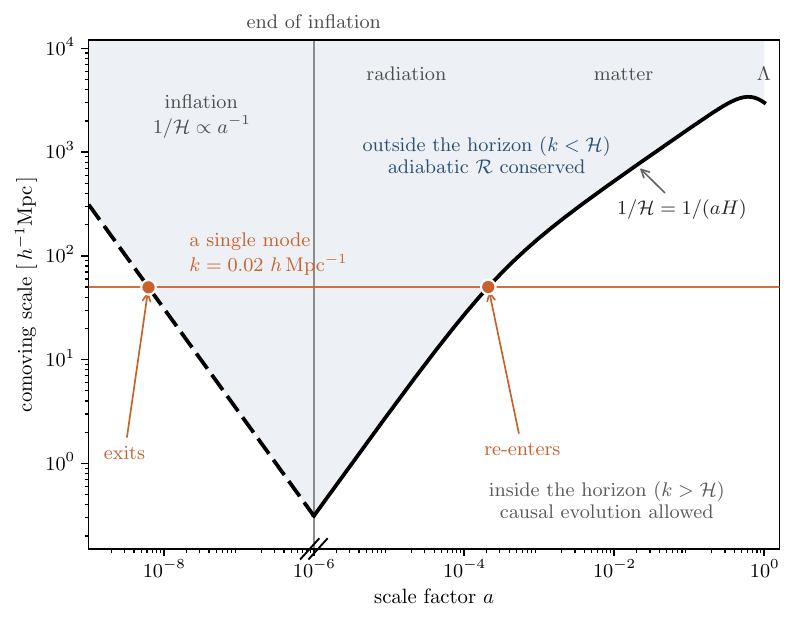}
\caption{Horizon exit and re-entry of a fixed comoving mode.
The thick curve is the comoving Hubble radius $1/\cH$ and the horizontal line is a single mode at fixed comoving $1/k$.
The post-inflationary branch is computed from the Friedmann equation used throughout these notes, giving $a_{\rm eq}=2.9\times10^{-4}$ and $k_{\rm eq}=\cH(a_{\rm eq})=0.0155\,h\,\mathrm{Mpc}^{-1}$;
its slopes on logarithmic axes are $+1$, $+1/2$ and $-1$ under radiation, matter and $\Lambda$.
The inflationary branch is schematic, and the break on the axis marks where the horizontal scale is compressed:
physical reheating lies approximately 20 orders of magnitude earlier in $a$ than shown.
The shaded region denotes the super-horizon regime of the adiabatic curvature mode.}
\label{fig:horizonhistory}
\end{figure}

Since $\vk$ is constant, a given mode is a horizontal line in \cref{fig:horizonhistory}, and its horizon status changes when $1/\cH$ crosses that line.
During inflation, $H$ is nearly constant and $a$ grows, so $1/\cH=1/(aH)$ shrinks and modes \emph{exit}, and afterwards $1/\cH$ grows again and the same modes \emph{re-enter}, the largest scales last.
For an adiabatic Fourier mode, $\mathcal R(\vk)$ is conserved outside the horizon after its decaying mode becomes negligible.
Its value at horizon re-entry is therefore the value set during inflation, making the primordial curvature field of \cref{sec:primordial-curvature} an initial condition for late-time evolution.
A mode that re-entered during radiation domination underwent less growth than one that re-entered later.
This scale-dependent suppression shapes $\Plin(k)$ in \cref{sec:linear-spectrum-shape}.

\subsection{The Newtonian fluid and linear growth}

Well inside the horizon and before shell crossing, clustering non-relativistic matter can be treated as a Newtonian, pressureless, single-stream and self-gravitating fluid.
The matter contains cold dark matter and baryons on scales where baryonic pressure is negligible.
In conformal time $\tau$, with peculiar velocity $\vu=\dd\vx/\dd\tau$, the nonlinear fluid equations are~\cite{MaBertschinger,DodelsonSchmidt}
\begin{align}
 \dot\delta+\nabla\cdot[(1+\delta)\vu]&=0,
 \label{eq:continuityfull}\\
 \dot\vu+\cH\vu+(\vu\cdot\nabla)\vu&=-\nabla\Phi,
 \label{eq:eulerfull}\\
 \nabla^2\Phi&=\frac32\Omega_m(\tau)\cH^2\delta,
 \label{eq:poisson}
\end{align}
which are respectively mass conservation, Newton's second law, and the field equation that makes the fluid self-gravitating.
We write $\theta\equiv\nabla\cdot\vu$ for the velocity divergence.
Taking the divergence of Euler before making any approximation gives
\begin{equation}
 \dot\theta+\cH\theta
 +\partial_i(u_j\partial_j u_i)=-\nabla^2\Phi.
 \label{eq:euler-divergence}
\end{equation}
Expand each perturbation in powers of the initial density field,
\begin{equation}
 \delta=\sum_{n\geq1}\delta^{(n)},
 \qquad
 \vu=\sum_{n\geq1}\vu^{(n)},
 \qquad
 \Phi=\sum_{n\geq1}\Phi^{(n)}.
 \label{eq:fluid-perturbative-series}
\end{equation}
The superscript $(n)$ denotes order $n$ in the initial density field.
The products $\delta\vu$ and $u_j\partial_j u_i$ begin at second order.
Retaining first-order terms and using Poisson gives the linear system
\begin{equation}
 \dot\delta=-\theta,
 \qquad
 \dot\theta+\cH\theta=-\frac32\Omega_m\cH^2\delta.
 \label{eq:linearsystem}
\end{equation}
The following elimination fixes the sign of the self-gravity term:
\begin{equation}
 \ddot\delta=-\dot\theta
 =\cH\theta+\frac32\Omega_m\cH^2\delta
 =-\cH\dot\delta+\frac32\Omega_m\cH^2\delta.
 \label{eq:growth-elimination}
\end{equation}
Moving everything to the left gives the growth equation
\begin{equation}
 \ddot\delta+\cH\dot\delta-\frac32\Omega_m\cH^2\delta=0.
 \label{eq:growth}
\end{equation}
We normalize the growing solution by its limit during matter domination and separately define a version normalized to unity today, as often used by numerical codes:
\begin{equation}
 \lim_{a\to0}\frac{D_+(a)}a=1,
 \qquad
 \widehat D_+(a)\equiv\frac{D_+(a)}{D_+(1)},
 \qquad
 \widehat D_+(1)=1.
 \label{eq:growth-normalization}
\end{equation}
Specified initial data evolve according to the ratio
\begin{equation}
 \delta(\vk,a)=\frac{D_+(a)}{D_+(a_{\rm ini})}\delta(\vk,a_{\rm ini}),
 \qquad
 f(a)\equiv\frac{\dd\ln D_+}{\dd\ln a}
 =\frac{\dd\ln\widehat D_+}{\dd\ln a}.
 \label{eq:growthfactor}
\end{equation}
In Einstein--de Sitter (EdS) we have $D_+\propto a$ and $f=1$, while in late-time flat $\Lambda$CDM $f\simeq\Omega_m^{0.55}$~\cite{Linder}.
Since $\dd\ln a/\dd\tau=\cH$, the definition of $f$ supplies the intermediate identity $\dot D_+=\cH fD_+$ (and the same one for $\widehat D_+$).
Continuity then gives
\begin{equation}
 \theta(\vk)=-\dot\delta(\vk)=-\cH f\,\delta(\vk),
 \label{eq:thetaf}
\end{equation}
which is the dynamical input behind the Kaiser redshift-space distortion we derive in \lectureref{3}.

\begin{warningbox}{assumptions of the linear fluid model}
Four assumptions are required for closure:
weak metric potentials, non-relativistic matter, sub-horizon modes, and a single-stream fluid.
Violations of the first three introduce gradual corrections.
Shell crossing invalidates the single-stream description abruptly because several velocities then occupy the same position;
\cref{sec:shell-crossing} develops this point.
\end{warningbox}

\subsection{Primordial curvature perturbations}
\label{sec:primordial-curvature}

Inflation supplies a nearly scale-invariant primordial curvature field $\mathcal R$~\cite{Planck2018}.
Here $\mathcal R$ is the comoving curvature perturbation, a dimensionless first-order perturbation of the spatial metric, $g_{ij}=a^2(1+2\mathcal R)\delta_{ij}$ on comoving slices.
At linear order, the three-dimensional Ricci scalar of a comoving spatial slice is
\[
 -4a^{-2}\nabla^2\mathcal R,
\]
or equivalently, $4k^2\mathcal R/a^2$ in Fourier space\footnote{Note that $\mathcal R$ should not be confused with the four-dimensional Ricci scalar in the Einstein equations, which carries dimensions of inverse length squared.}.
Because $\mathcal R$ is conserved outside the horizon for adiabatic perturbations, it provides the inflationary initial condition at horizon re-entry.
We parameterize its spectrum by the amplitude $A_s$ and tilt $n_s-1$ about the pivot scale $k_*$:
\begin{equation}
 \Delta_{\mathcal R}^2(k)
 \equiv\frac{k^3P_{\mathcal R}(k)}{2\pi^2}
 =A_s\left(\frac{k}{k_*}\right)^{n_s-1},
 \label{eq:primordial}
\end{equation}
with $n_s=1$ exactly scale invariant;
observations find a small red tilt, $n_s<1$~\cite{Planck2018}.
Gravity and the cosmic matter content process each primordial mode through a transfer function~\cite{EisensteinHu,CLASS}.

\begin{readingbox}{where $\mathcal R$ comes from}
Slow-roll inflation is driven by a scalar field $\phi$ rolling down a flat potential.
The near-de~Sitter expansion stretches its quantum fluctuations until they freeze outside the horizon.
For a light canonical field in quasi-de~Sitter space, define $v\equiv a\delta\phi$.
The Bunch--Davies solution to its leading mode equation is~\cite{BaumannCosmology}
\[
 v_k''+\left(k^2-\frac{2}{\tau^2}\right)v_k=0,
 \qquad
 v_k=\frac{e^{-\ii k\tau}}{\sqrt{2k}}
 \left(1-\frac{\ii}{k\tau}\right),
\]
where a prime denotes $\dd/\dd\tau$.
Taking the superhorizon limit $|k\tau|\ll1$ and using $a=-1/(H\tau)$ gives
\[
 |\delta\phi_k|^2=\left|\frac{v_k}{a}\right|^2
 \longrightarrow\frac{H^2}{2k^3},
 \qquad
 \Delta_{\delta\phi}^2
 \equiv\frac{k^3}{2\pi^2}|\delta\phi_k|^2
 =\left(\frac{H}{2\pi}\right)^2.
\]
On spatially flat slices, the metric convention $g_{ij}=a^2(1+2\mathcal R)\delta_{ij}$ gives
\[
 \mathcal R=-\frac{H}{\dot\phi}\,\delta\phi.
\]
Define the potential slow-roll parameters by
\[
 \epsilon_V\equiv\frac{M_{\rm Pl}^2}{2}
 \left(\frac{V_{,\phi}}V\right)^2,
 \qquad
 \eta_V\equiv M_{\rm Pl}^2\frac{V_{,\phi\phi}}V.
\]
The subscripts distinguish them from the stochastic field $\epsilon$ of \lectureref{2} and the FFTLog frequency $\eta_m$ defined in Appendix~\ref{app:master-integrals}.
The leading slow-roll background equations give
\[
 3H\dot\phi\simeq-V_{,\phi},
 \qquad 3M_{\rm Pl}^2H^2\simeq V
 \quad\Longrightarrow\quad
 \dot\phi^2\simeq2\epsilon_VM_{\rm Pl}^2H^2.
\]
Substituting the field spectrum then gives
\[
 \Delta_{\mathcal R}^2
 =\left(\frac{H}{\dot\phi}\right)^2\Delta_{\delta\phi}^2
 =\frac{H^2}{8\pi^2M_{\rm Pl}^2\epsilon_V}
 \simeq\frac{V}{24\pi^2M_{\rm Pl}^4\epsilon_V}
 \bigg|_{k=aH}.
\]
At horizon exit, $\dd/\dd\ln k\simeq-M_{\rm Pl}^2(V_{,\phi}/V)\dd/\dd\phi$.
Differentiating at horizon exit gives
\[
 \frac{\dd\ln V}{\dd\ln k}\simeq-2\epsilon_V,
 \qquad
 \frac{\dd\ln\epsilon_V}{\dd\ln k}\simeq4\epsilon_V-2\eta_V,
 \qquad
 n_s-1\equiv\frac{\dd\ln\Delta_{\mathcal R}^2}{\dd\ln k}
 =-6\epsilon_V+2\eta_V.
\]
Thus the amplitude $A_s$ of \cref{eq:primordial} measures $H^2/\epsilon_V$ when each scale left the horizon, while slow evolution produces the small tilt.
Single-field slow roll also predicts primordial non-Gaussianity suppressed by the slow-roll parameters~\cite{BaumannCosmology}, supporting the Gaussian initial conditions used in \lectureref{3} and the \extensionref.
\end{readingbox}

The transfer function maps the primordial curvature statistics to the late-time matter field.
With the early-time normalization of \cref{eq:growth-normalization}, the potential relation during matter domination and the Poisson equation fix its sign and amplitude.
\begin{derivationbox}{from curvature to density}
For the adiabatic growing mode, the standard matter-dominated era potential transfer relation~\cite{DodelsonSchmidt,BaumannCosmology} and Poisson equation give
\begin{equation}
 \Phi(\vk,a)=-\frac35\frac{D_+(a)}aT(k)\mathcal R(\vk),
 \qquad
 -k^2\Phi=\frac32\Omega_m\cH^2\delta_m.
 \label{eq:curvature-to-density}
\end{equation}
The background Friedmann equation gives $\Omega_m(a)\cH^2(a)=\Omega_{m0}H_0^2/a$.
Substituting this identity into Poisson produces two minus signs, one from Poisson and one from the potential--curvature relation, and they cancel:
\[
 \delta_m
 =-\frac{2k^2\Phi}{3\Omega_m\cH^2}
 =\frac25\frac{k^2T(k)D_+(a)}{\Omega_{m0}H_0^2}\mathcal R.
\]
The positive transfer kernel below is the resulting master relation.
If one replaces $D_+$ by $\widehat D_+$, then $D_+(a)=D_+(1)\widehat D_+(a)$ and the constant $D_+(1)$ must remain in the kernel;
omitting it changes the amplitude normalization.
\end{derivationbox}
Thus the matter fluctuation is linear in the curvature field,
\begin{equation}
 \delta_m(\vk,a)=\mathcal M(k,a)\mathcal R(\vk),
 \qquad
 \mathcal M(k,a)=\frac{2}{5}\frac{k^2T(k)D_+(a)}{\Omega_{m0}H_0^2}.
 \label{eq:Mtransfer}
\end{equation}
The linear matter power spectrum is the primordial one processed by the square of that kernel,
\begin{equation}
 \Plin(k,a)=\mathcal M^2(k,a)P_{\mathcal R}(k).
 \label{eq:PLfromR}
\end{equation}
The factorization is independent of whether growth is normalized by $D_+(a)/a\to1$ at early times or by $\widehat D_+(1)=1$ today.
Primordial amplitude and tilt, transfer through the radiation and matter eras, and late-time growth remain three separable contributions.
At fixed redshift, combining \cref{eq:primordial,eq:PLfromR} gives $\Plin(k)\propto\Delta_{\mathcal R}^2(k)\,k\,T^2(k)$.
The single factor $k$ is what remains of the $k^4$ curvature-to-density conversion after the $k^{-3}$ carried by $P_{\mathcal R}(k)=2\pi^2\Delta_{\mathcal R}^2(k)/k^3$ is made explicit.

\subsection{Reading the shape of the linear matter power spectrum}
\label{sec:linear-spectrum-shape}

\begin{figure}[t]
\centering
\includegraphics[width=0.96\textwidth]{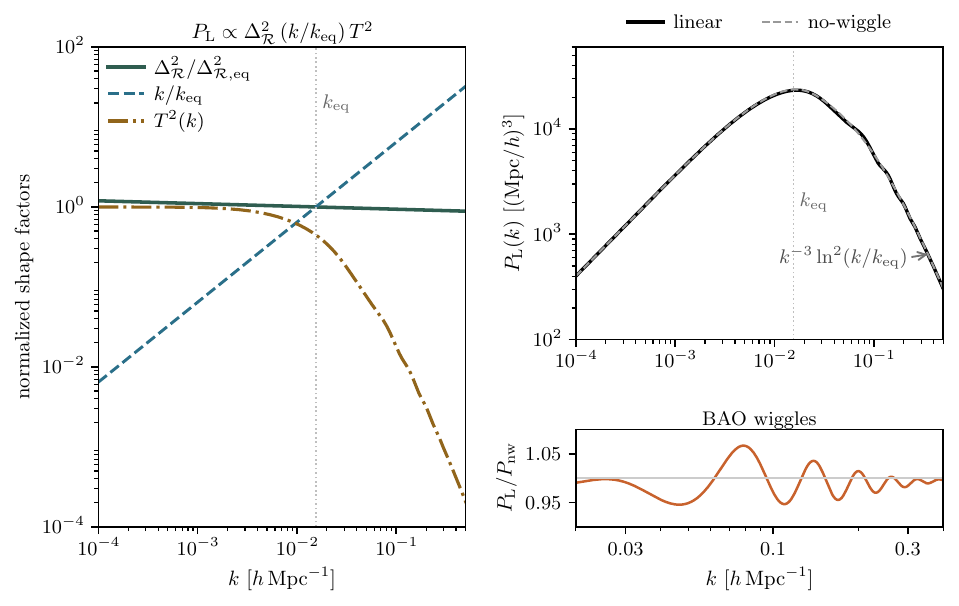}
\caption{Construction of the linear matter power spectrum.
\emph{Left:} three dimensionless shape factors $\Delta_{\mathcal R}^2(k)/\Delta_{\mathcal R}^2(k_{\rm eq})\propto (k/k_{\rm eq})^{n_s-1}$, $k/k_{\rm eq}$, and $T^2(k)$, where the transfer function is normalized by its large-scale limit, $T(k\to0)=1$.
The dashed blue curve is the residual $k$ left by the $k^4$ curvature-to-density conversion after writing $P_{\mathcal R}\propto\Delta_{\mathcal R}^2/k^3$;
the product of all three factors is proportional to $\Plin(k)$.
\emph{Right:} the dimensional linear matter power spectrum $\Plin(k)$ with a baryon-free power spectrum overlaid.
This curve is labelled no-wiggle in the legend and denoted by $P_{\rm nw}$ throughout the notes;
the lower-right subpanel isolates the baryon acoustic oscillations.
The transfer suppression at the turnover is $T^2(k_{\rm eq})=0.447$ for the cosmology used throughout these notes, and the high-$k$ tail approaches $k^{n_s-4}\ln^2(k/k_{\rm eq})\simeq k^{-3}\ln^2(k/k_{\rm eq})$.}
\label{fig:pkconstruction}
\end{figure}

\cref{fig:pkconstruction} organizes the construction of $\Plin(k)$ into three physical features.
The vertical axis ranges differ between panels, so their slopes should not be compared by eye.
At fixed redshift, $\Plin\propto k^{n_s}T^2$, and hence
\begin{equation}
 \frac{d\ln\Plin}{d\ln k}=n_s+\frac{d\ln T^2}{d\ln k}.
 \label{eq:transfer-slope-check}
\end{equation}
At high $k$, $T^2\propto k^{-4}\ln^2(k/k_{\rm eq})$, so $\Plin\propto k^{n_s-4}\ln^2(k/k_{\rm eq})$.
For $n_s\simeq1$, $T^2$ therefore falls by approximately one more power of $k$ than $\Plin$.
\begin{enumerate}
\item On sufficiently large scales, modes entered the horizon after matter--radiation equality and retain the primordial slope.
\item Modes that entered during radiation domination could not grow efficiently.
This is the M\'esz\'aros effect, and the suppression it leaves creates the turnover near the equality scale and the falling high-$k$ tail~\cite{Meszaros,DodelsonSchmidt}.
\item Before recombination, pressure in the tightly coupled photon--baryon fluid provided the restoring force for acoustic oscillations.
Their fossil imprint survives as the baryon acoustic oscillation (BAO) feature:
wiggles in $\Plin(k)$ and a bump in the two-point correlation function~\cite{PeeblesYu,EisensteinBAO}.
\end{enumerate}
\paragraph{Why radiation stalls growth.}
The turnover scale follows from a competition between two clocks.
A perturbation collapses on the free-fall time of the matter it contains, $t_{\rm ff}\sim(G\bar\rho_m)^{-1/2}$, while the background stretches it apart on the expansion time $t_{\rm exp}\sim H^{-1}$.
In matter domination, $H^2\sim G\bar\rho_m$, so the two timescales are comparable and $\delta$ can grow steadily as $D_+\propto a$.
During radiation domination the radiation dominates the expansion but does not cluster, so $H^2\sim G\bar\rho_r\gg G\bar\rho_m$ and the expansion clock runs much faster than the free-fall clock.
The matter perturbation then cannot grow appreciably under its own gravity and evolves only logarithmically rather than as a power of $a$.
A mode that entered the horizon earlier, and therefore has higher $k$, remains in this suppressed regime for longer.
This logarithmic growth is the M\'esz\'aros effect.
The transfer function $T(k)$ records the growth lost before matter domination.
We normalize $T(k\to0)=1$ for scales that entered late;
modes that entered early then acquire the suppression $T(k)\propto k^{-2}$ up to logarithms.

At linear order these features have fixed shape, and all subsequent redshift dependence enters through the single factor $D_+^2(z)$:
\begin{equation}
 \Plin(k,z)=\left[\frac{D_+(z)}{D_+(z_0)}\right]^2\Plin(k,z_0).
 \label{eq:PLgrowth}
\end{equation}

Before defining the amplitude $\sigma_8$, consider the smoothing operation.
For a normalized spherical window $W_R$, convolution in position space becomes multiplication in Fourier space, $\delta_R(\vk)=W(kR)\delta(\vk)$.
Applying \cref{eq:powerdef} gives
\begin{equation}
 \sigma_R^2\equiv\langle\delta_R^2(\vx)\rangle
 =\int\frac{\dd^3k}{(2\pi)^3}P(k)|W(kR)|^2
 =\int\dd\ln k\,\Delta^2(k)|W(kR)|^2.
 \label{eq:smoothing-variance}
\end{equation}
Smoothing the linear field with a spherical top-hat of radius $8\,h^{-1}\mathrm{Mpc}$ then gives
\begin{equation}
 \sigma_8^2(z)=\int\dd\ln k\,\Delta^2(k,z)\,|W(kR_8)|^2,
 \qquad R_8=8\,h^{-1}\mathrm{Mpc},
 \label{eq:sigma8}
\end{equation}
with $\Delta^2$ built from $\Plin$ as in \cref{eq:deltadimless}, and where $W$ is the Fourier transform of a real-space top-hat, $W(x)=3(\sin x-x\cos x)/x^3$, normalized to $W(0)=1$.
Because $\Plin\propto\sigma_8^2$ at fixed shape, $\sigma_8$ parameterizes the amplitude of the linear matter power spectrum.
This dependence leads to the combinations $b_1\sigma_8$ and $f\sigma_8$ measured in \lectureref{3}.

\begin{keybox}{Lecture 1---four statements to retain}
The density contrast is the field;
Fourier modes provide its scale decomposition, and $P(k)$ is their variance.
The linear matter power spectrum $\Plin(k,z)$ combines primordial statistics, transfer physics and linear growth.
The same $D_+$ that evolves the density field also fixes the velocity field through $f=\dd\ln D_+/\dd\ln a$, relating both fields to a single function of time.
\end{keybox}

\phantomsection
\label{exercises:lecture1}
\subsection*{Exercises}
\begin{enumerate}
\item Specialize \cref{eq:growth} to EdS, where $a\propto\tau^2$, $\cH=2/\tau$, and $\Omega_m=1$.
Find both power-law solutions and express them as powers of $a$.
\item Starting from $\delta_R(\vk)=W(kR)\delta(\vk)$, use \cref{eq:powerdef} to derive \cref{eq:smoothing-variance}.
Identify where reality and spherical symmetry of the window enter.
\item A code implements
\[
 \mathcal M_{\rm code}(k,a)
 =-\frac25\frac{k^2T(k)\widehat D_+(a)}{\Omega_{m0}H_0^2}.
\]
Audit its dimensions, sign, and normalization against \cref{eq:curvature-to-density,eq:Mtransfer}.
Which error is invisible in $\Plin=\mathcal M^2P_{\mathcal R}$ but visible in $\delta_m/\mathcal R$?
\end{enumerate}

\lecturetitle{2}{Gravitational collapse and the cosmic web}

\begin{objectives}
This lecture describes nonlinear gravitational evolution.
Students should be able to
\begin{itemize}
\item derive the idealized spherical-collapse model;
\item formulate the Lagrangian description of cosmological fluids and the anisotropic collapse of the Zel'dovich approximation;
\item define shell crossing and identify the assumptions it violates;
\item obtain linear galaxy bias as the response of a thresholded population to a long-wavelength density perturbation.
\end{itemize}
\end{objectives}

\subsection{Why linear growth is not a theory of collapse}

\cref{eq:growth} predicts $\delta_{\rm L}\propto D_+$ without limit and therefore fails once $|\delta|\sim1$, when the terms omitted from \cref{eq:continuityfull,eq:eulerfull} are no longer small.
Linear theory nevertheless provides a useful clock:
extrapolating the initial fluctuation with $D_+$ associates each nonlinear event with a value of $\delta_{\rm L}$.
Spherical collapse gives this correspondence exactly under the restrictive assumption of spherical symmetry.

\subsection{Spherical collapse: turnaround, formal collapse, and virialization}

\begin{figure}[htbp]
\centering
\begin{cmfigurefont}
\begin{tikzpicture}[>=Latex,scale=0.95,every node/.style={font=\small}]
\draw[->,Ink!65] (0,0)--(12,0) node[right]{time};
\foreach \x/\lab in {0.8/linear,4.0/turnaround,7.2/formal collapse,10.6/virialized halo}
 {\draw[Ink!55] (\x,-0.12)--(\x,0.12) node[below=5pt,align=center]{\lab};}
\draw[Ink!30,dashed] (0.7,2.0)--(7.9,2.0)
 node[right,color=Ink!70]{$R_{\rm ta}$};
\draw[DeepGreen!35,dashed] (0.7,1.0)--(7.2,1.0);
\draw[very thick,LinkBlue,domain=0:360,samples=160,smooth]
 plot ({0.8+1.0186*(\x*pi/180-sin(\x))},{1-cos(\x)});
\draw[very thick,DeepGreen] (7.2,1.0)--(10.9,1.0)
 node[right,color=DeepGreen]{$R_{\rm ta}/2$};
\node[color=LinkBlue] at (1.25,1.58) {$R(t)$};
\node[color=DeepGreen!85] at (9.1,1.34) {virialization};
\end{tikzpicture}
\end{cmfigurefont}
\caption{Spherical collapse and virialization.
The cycloid of \cref{eq:cycloid} reaches zero radius at formal collapse.
Virialization is a separate physical prescription representing orbiting matter and leaves the halo at half the turnaround radius.}
\label{fig:cycloid}
\end{figure}
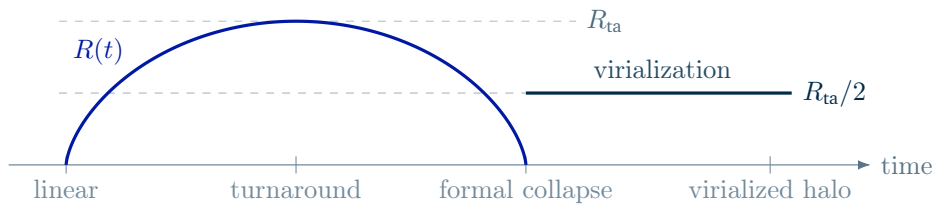

We consider a uniform spherical top-hat of fixed mass $M$ and physical radius $R(t)$ in an EdS background~\cite{GunnGott}.
Matter outside the sphere exerts no net force on it.
In general relativity Birkhoff's theorem gives the same result, so the boundary obeys
\begin{equation}
 \ddot R=-\frac{GM}{R^2},
 \label{eq:sphereR}
\end{equation}
whose first integral is a specific energy $E=\tfrac12\dot R^2-GM/R$.
At an early time $t_i$, the velocity must contain the peculiar infall of the pure growing mode, rather than being set equal to the Hubble velocity~\cite{Jenkins2010}.
Mass conservation and linear continuity give
\begin{equation}
 M=\frac{4\pi}{3}\bar\rho_i(1+\delta_i)R_i^3,
 \qquad
 \dot R_i=H_iR_i\left(1-\frac{\delta_i}{3}\right)+\Order(\delta_i^2).
 \label{eq:growing-mode-initialization}
\end{equation}
Defining $K_{H,i}\equiv\tfrac12(H_iR_i)^2$, the EdS Friedmann equation implies $GM/R_i=K_{H,i}(1+\delta_i)$.
The corrected kinetic energy is $K_{H,i}(1-2\delta_i/3)$ at first order, and hence
\begin{equation}
 E_i=-\frac53K_{H,i}\delta_i+\Order(\delta_i^2).
 \label{eq:spherical-energy}
\end{equation}
An overdense growing-mode patch is therefore bound.
Integrating its negative-energy orbit as shown in \cref{app:spherical-collapse} gives the universal cycloid
\begin{equation}
 R(\eta)=A(1-\cos\eta),
 \qquad
 t(\eta)=B(\eta-\sin\eta),
 \qquad A^3=GMB^2.
 \label{eq:cycloid}
\end{equation}
The patch expands more slowly than the background, turns around at $\eta=\pi$, and formally collapses to a point at $\eta=2\pi$.
The top-hat and background densities are
\begin{equation}
 \rho_{\rm th}=\frac{3M}{4\pi R^3},
 \qquad
 \bar\rho_{\rm EdS}(t)=\frac{1}{6\pi Gt^2}.
 \label{eq:spherical-densities}
\end{equation}
Inserting the cycloid into their ratio gives
\begin{equation}
 1+\delta_{\rm NL}(\eta)
 =\frac{9}{2}\frac{(\eta-\sin\eta)^2}{(1-\cos\eta)^3}.
 \label{eq:deltanlsphere}
\end{equation}
At turnaround this ratio is $9\pi^2/16\simeq5.6$;
at formal collapse it diverges.
Matching the early-time nonlinear solution to the EdS growing mode, as shown in \cref{app:spherical-collapse}, gives
\begin{equation}
 \delta_{\rm L}(\eta)
 =\frac35\left[\frac34(\eta-\sin\eta)\right]^{2/3}.
 \label{eq:deltalsphere}
\end{equation}
At formal collapse,
\begin{equation}
 \delta_c\equiv\delta_{\rm L}(2\pi)
 =\frac{3}{20}(12\pi)^{2/3}\simeq1.686.
 \label{eq:deltac}
\end{equation}
The same calculation yields the $17/21$ perturbative check in \cref{eq:sphereperturbative}.
The threshold $\delta_c$ is a linearly extrapolated value, not the actual nonlinear density.
An EdS universe carries no preferred scale, so $\delta_c$ is the same for a collapsing region of any mass, and we use it as a single threshold for haloes of every size in the galaxy-bias construction of \cref{sec:collapsed-to-biased-tracers}.
The virial theorem~\cite{DodelsonSchmidt}, $K=-U/2$, leaves the relaxed object at half its turnaround radius, and \cref{fig:cycloid} shows these three stages on one timeline.
Halving the radius multiplies its density by eight.
Between turnaround and collapse, the background density decreases by another factor of four.
The virialized overdensity is therefore $\Delta_{\rm vir}=(9\pi^2/16)\times8\times4=18\pi^2\simeq178$.
With $\Lambda$ present, the equation of motion has no closed-form solution and both characteristic values shift mildly~\cite{BryanNorman}.
Collapse typically completes during matter domination, before the region is affected strongly by $\Lambda$.
Both values also depend on the reference density used to define them.

\begin{warningbox}{spherical-collapse idealization}
Spherical symmetry is an idealized setup.
Exact sphericity requires three equal deformation eigenvalues and is therefore a measure-zero condition for a Gaussian random field.
Generic patches collapse anisotropically, as the Zel'dovich approximation makes explicit.
\end{warningbox}

\subsection{The Lagrangian map and mass conservation}

\begin{figure}[t]
\centering
\includegraphics[width=0.96\textwidth]{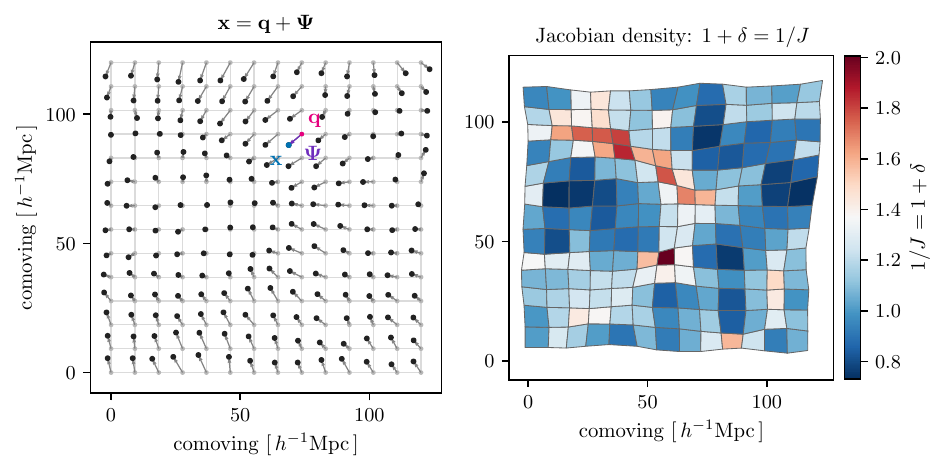}
\caption{Lagrangian displacement and the Jacobian.
\emph{Left:} a regular grid of Lagrangian coordinates $\vq$, with arrows to the positions reached under the Zel'dovich displacement.
The coordinates remain fixed labels while the matter elements move.
\emph{Right:} the displaced cells drawn as quadrilaterals and shaded by their inverse area ratio, which in two dimensions is $1/J=1+\delta$.
Contracting cells form an overdense ridge, marking the onset of a filament.
The illustrative displacement amplitude is $0.46$ of a cell, giving $1/J$ between $0.73$ and $2.00$.}
\label{fig:lagrangianmap}
\end{figure}

As illustrated in \cref{fig:lagrangianmap}, the Eulerian description specifies fields at fixed locations $\vx$, whereas the Lagrangian description labels each matter element by its initial coordinate $\vq$ and follows its displacement $\vPsi$,
\begin{equation}
 \vx(\vq,\tau)=\vq+\vPsi(\vq,\tau).
 \label{eq:lptmap}
\end{equation}
We expand the displacement rather than the density,
\begin{equation}
 \vPsi=\sum_{n\geq1}\vPsi^{(n)}
 =\vPsi^{(1)}+\vPsi^{(2)}+\cdots,
 \label{eq:lpt-series}
\end{equation}
using the order convention of \cref{eq:fluid-perturbative-series}.
Mass conservation then reduces to a relation between volumes.
A small initial volume $\dd^3q$ becomes $\dd^3x=J\dd^3q$, where the Jacobian of the map is
\begin{equation}
 J(\vq,\tau)=\det\!\left(\delta_{ij}+\Psi_{i,j}\right),
 \qquad
 1+\delta[\vx(\vq),\tau]=\frac{1}{J(\vq,\tau)}.
 \label{eq:lptjac}
\end{equation}
A cell that shrinks has $J<1$ and is therefore overdense, and the relation remains exact for as long as the map remains single-valued.
At first order, $J=1+\nabla_q\cdot\vPsi^{(1)}+\Order(\Psi^2)$, so
\begin{equation}
 \nabla_q\cdot\vPsi^{(1)}=-\delta^{(1)},
 \qquad
 \vPsi^{(1)}(\vk)=\frac{\ii\vk}{k^2}\delta^{(1)}(\vk),
 \label{eq:psilinear}
\end{equation}
where the second expression assumes the growing, irrotational solution.
The dynamics supplies the same time dependence as for the linear density, $\vPsi^{(1)}\propto D_+$.

\FloatBarrier
\subsection{Anisotropic collapse: the Zel'dovich approximation and the cosmic web}

The Zel'dovich approximation (ZA)~\cite{Zeldovich} inserts the first-order displacement into the full nonlinear map.
Particles then follow straight comoving trajectories, while the density is computed from the exact determinant.
We write $\vPsi^{(1)}(\vq,a)=D_+(a)\vPsi_0(\vq)$ and define the symmetric deformation tensor and its principal axes by
\begin{equation}
 d_{ij}\equiv-\Psi_{0i,j},
 \qquad
 O^{\mathsf T}dO=\operatorname{diag}(\lambda_1,\lambda_2,\lambda_3).
 \label{eq:za-deformation}
\end{equation}
The Jacobian matrix is therefore $I-D_+d$, so \cref{eq:lptjac} becomes
\begin{equation}
 J_{\rm ZA}=\det(I-D_+d)=\prod_{i=1}^{3}(1-D_+\lambda_i),
 \qquad
 1+\delta_{\rm ZA}=\prod_{i=1}^{3}(1-D_+\lambda_i)^{-1}.
 \label{eq:zadensity}
\end{equation}
ZA truncates the displacement, not its determinant:
the displacement here is only first order, but the determinant retains every power of it.
The ZA therefore retains bulk displacements that a fixed-order Eulerian density expansion truncates.

\begin{figure}[t]
\centering
\includegraphics[width=0.96\textwidth]{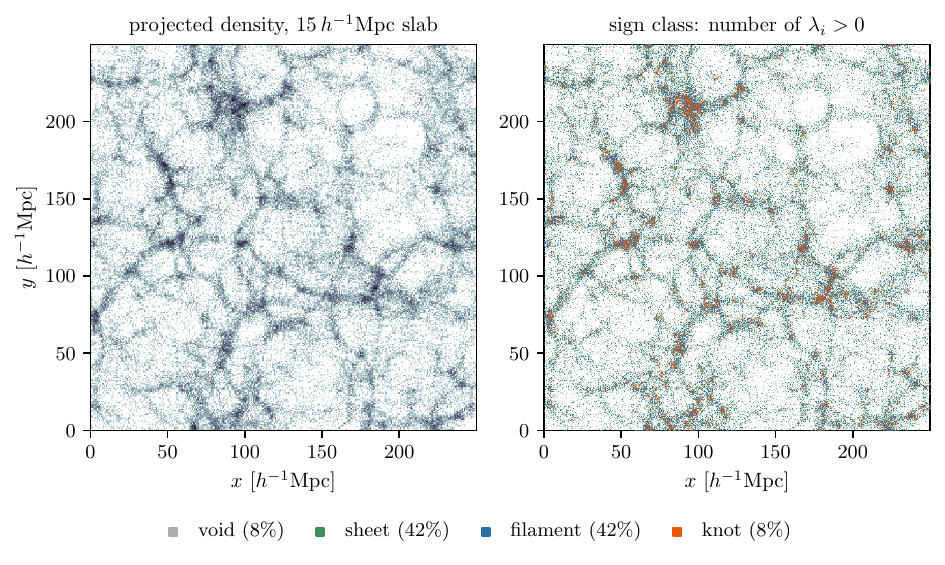}
\caption{Cosmic-web classification from deformation eigenvalues.
\emph{Left:} the projected density of a Zel'dovich-displaced field.
\emph{Right:} particles sampled uniformly in Lagrangian space and classified by the number of positive deformation eigenvalues.
The plotted realization gives initial Lagrangian volume fractions of $8.0$, $42.1$, $42.0$ and $8.0$ per cent for voids, sheets, filaments and knots, respectively, close to the $8/42/42/8$ per cent predicted by the Doroshkevich distribution for the eigenvalues of a Gaussian deformation tensor~\cite{Doroshkevich,BiasReview}.
This classification by the signs of the eigenvalues identifies which directions contract as $D_+$ grows;
it does not assert completed collapse at the plotted epoch.
The right panel plots displaced positions, so its projected area fractions differ from these initial Lagrangian fractions because the Zel'dovich map has redistributed mass from voids to knots.}
\label{fig:cosmicweb}
\end{figure}

The eigenvalue structure of this determinant also explains why generic collapse produces a web rather than a collection of spherical objects.
\cref{fig:cosmicweb} displays both the displaced density and its classification by deformation eigenvalues.

\begin{figure}[t]
\centering
\begin{cmfigurefont}
\begin{tikzpicture}[>=Latex,node distance=22mm,every node/.style={font=\small}]
\tikzset{
 stagebox/.style={draw=Ink!60,fill=SoftBlue,minimum height=12mm,
 align=center,inner xsep=3pt},
 threshold/.style={above=2pt,color=LinkBlue,
 fill=white,inner sep=1pt}}
\node[stagebox,minimum width=17mm] (cube) {patch};
\node[stagebox,right=of cube,minimum width=25mm,minimum height=5mm] (sheet) {sheet};
\node[stagebox,right=of sheet,minimum width=28mm,minimum height=3mm] (fil) {filament};
\node[stagebox,right=of fil,minimum width=12mm,minimum height=8mm] (knot) {knot};
\node[anchor=south] at ($(cube.west)!0.5!(knot.east)+(0,12mm)$)
 {conditional all-positive path: $\lambda_1>\lambda_2>\lambda_3>0$};
\draw[->,thick,LinkBlue] (cube)--node[threshold]{$D_+\lambda_1=1$}(sheet);
\draw[->,thick,LinkBlue] (sheet)--node[threshold]{$D_+\lambda_2=1$}(fil);
\draw[->,thick,LinkBlue] (fil)--node[threshold]{$D_+\lambda_3=1$}(knot);
\end{tikzpicture}
\end{cmfigurefont}
\caption{Conditional collapse sequence for three positive eigenvalues.
Each arrow marks $D_+\lambda_i=1$, equivalently the vanishing of the Jacobian factor $1-D_+\lambda_i$.
Patches with zero, one or two positive eigenvalues do not traverse all three arrows.
Along the sequence shown, only the first crossing lies within the dynamical validity of the ZA.}
\label{fig:eigenvaluepath}
\end{figure}
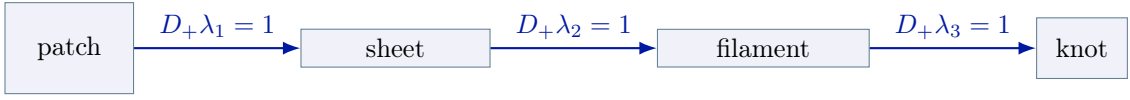

For a spherical patch $\lambda_1=\lambda_2=\lambda_3$, but for a generic random symmetric matrix the three eigenvalues are distinct.
A factor in \cref{eq:zadensity} reaches zero at finite positive $D_+$ only when its eigenvalue has $\lambda_i>0$, at $D_+=1/\lambda_i$;
a non-positive eigenvalue never crosses.
If all three are positive and we order them $\lambda_1>\lambda_2>\lambda_3>0$, their formal crossings occur in that order, which is the sequence shown in \cref{fig:eigenvaluepath}.
Along this conditional path, collapse therefore happens first along a single axis and produces a sheet (a.k.a.
Zel'dovich pancake);
a second positive eigenvalue then formally turns the sheet into a filament, and a third formally identifies a knot.
The resulting web is a generic consequence of anisotropic collapse rather than an additional structure imposed around spherical halos~\cite{Doroshkevich,ShandarinZeldovich}.

\FloatBarrier
\subsection{Shell crossing and the limit of the theory}
\label{sec:shell-crossing}

As the first factor in \cref{eq:zadensity} reaches zero, $J\to0$ and the density formally diverges.
The map $\vq\mapsto\vx$ then ceases to be invertible because trajectories with different velocities reach the same position.
This \emph{shell-crossing} event invalidates both the single-stream fluid description of \lectureref{1} and the interpretation of \cref{eq:lptjac} as a one-to-one map.
The sequence sheet $\to$ filament $\to$ knot remains a morphological guide.
A Zel'dovich knot is not a bound halo.
Binding, violent relaxation, and virialization require phase-space dynamics or an $N$-body simulation~\cite{ShandarinZeldovich}.

\begin{readingbox}{second-order LPT}
At first and second order the growing displacement is longitudinal.
Define the corresponding Lagrangian displacement potentials by
\[
 \vPsi^{(1)}=-D_+\nabla_q\phi^{(1)},
 \qquad
 \vPsi^{(2)}=D_2\nabla_q\phi^{(2)}.
\]
Equation~\eqref{eq:psilinear} then fixes $\nabla_q^2\phi^{(1)}=\delta^{(1)}/D_+$.
Define the map matrix and its cofactor matrix by
\[
 A_{ij}\equiv\frac{\partial x_i}{\partial q_j}
 =\delta_{ij}+\Psi_{i,j},
 \qquad
 C_{ij}\equiv J(A^{-1})_{ji}.
\]
Along $\vx(\vq,\tau)$, the convective derivative in \cref{eq:eulerfull} gives $\ddot\vPsi+\cH\dot\vPsi=-\nabla_x\Phi$.
Taking the Eulerian divergence, multiplying by $J$, and using Poisson together with $J(1+\delta)=1$ gives
\begin{equation}
 C_{ij}\left(\ddot\Psi_{i,j}+\cH\dot\Psi_{i,j}\right)
 =\frac32\Omega_m\cH^2(J-1).
 \label{eq:lpt-cofactor-eom}
\end{equation}
Let $B_{ij}\equiv\Psi^{(1)}_{i,j}$.
The quadratic expansions are
\[
 \begin{split}
 J={}&1+\operatorname{tr}B+\nabla_q\cdot\vPsi^{(2)}+I_2(B),
 \qquad
 C_{ij}=\delta_{ij}+\delta_{ij}\operatorname{tr}B-B_{ji},\\
 I_2(B)\equiv{}&\frac12\left[(\operatorname{tr}B)^2-\operatorname{tr}(B^2)\right].
 \end{split}
\]
Here $I_2$ is the second principal invariant of the first-order deformation tensor.
The first-order growth equation implies $\ddot\Psi^{(1)}_{i,j}+\cH\dot\Psi^{(1)}_{i,j} =(3/2)\Omega_m\cH^2B_{ij}$.
At second order, the cofactor equation reduces to
\begin{equation}
 \left(\partial_\tau^2+\cH\partial_\tau-\frac32\Omega_m\cH^2\right)
 \nabla_q\cdot\vPsi^{(2)}
 =-\frac32\Omega_m\cH^2 I_2(B).
 \label{eq:2lpt-longitudinal}
\end{equation}
Since $B_{ij}=-D_+\phi^{(1)}_{,ij}$ and $\nabla_q\cdot\vPsi^{(2)}=D_2\nabla_q^2\phi^{(2)}$, we place the full time-dependent coefficient in $D_2$ and normalize the spatial source to unit coefficient.
Separating \cref{eq:2lpt-longitudinal} then gives~\cite{Bouchet}
\begin{equation}
 \nabla_q^2\phi^{(2)}=
 \frac12\left[(\nabla_q^2\phi^{(1)})^2
 -\phi^{(1)}_{,ij}\phi^{(1)}_{,ij}\right]
 =
 \sum_{i>j}\left(\phi^{(1)}_{,ii}\phi^{(1)}_{,jj}
 -[\phi^{(1)}_{,ij}]^2\right).
 \label{eq:2lpt-source}
\end{equation}
The time-dependent coefficient obeys
\begin{equation}
 \ddot D_2+\cH\dot D_2-\frac32\Omega_m\cH^2D_2
 =-\frac32\Omega_m\cH^2D_+^2.
 \label{eq:D2-growth}
\end{equation}
In EdS, $D_+\propto\tau^2$ and $\cH=2/\tau$.
Substituting $D_2=cD_+^2$ into \cref{eq:D2-growth} gives $14c=-6$, hence $D_2=-3D_+^2/7$.
The ZA supplies straight streams;
2LPT supplies their leading tidal bending and strongly reduces transients in simulation initial conditions~\cite{Crocce}.
\end{readingbox}

\subsection{From collapsed regions to biased tracers}
\label{sec:collapsed-to-biased-tracers}

\begin{figure}[t]
\centering
\includegraphics[width=0.96\textwidth]{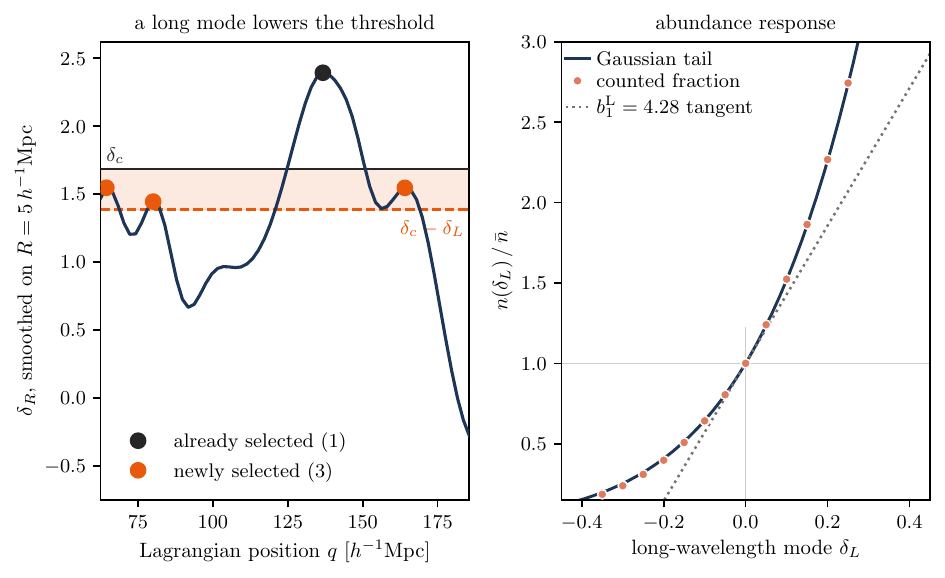}
\caption{Threshold bias as an abundance response.
\emph{Left:} the initial field smoothed on $R=5\,h^{-1}\mathrm{Mpc}$, with the collapse threshold $\delta_c$ drawn as the solid line.
A long mode $\delta_L$ leaves the field unchanged but lowers the cut to $\delta_c-\delta_L$, causing three previously sub-threshold peaks to cross it.
\emph{Right:} the abundance $F(\delta_L)$ of \cref{eq:threshold-tail}, with the tangent at the origin giving the Lagrangian response $b_1^{\rm L}$ of \cref{eq:thresholdbias}.
The large count response to an unchanged small-scale field is the threshold-bias mechanism.}
\label{fig:thresholdbias}
\end{figure}

Galaxy surveys count galaxies rather than mass elements.
Because galaxy formation selects particular environments, the galaxy density contrast need not equal the matter density contrast.
The threshold model below illustrates the response principle~\cite{PressSchechter,Kaiser1984,BBKS}.
Ref.~\cite{BiasReview} develops the systematic bias expansion and its renormalization, and the \extensionref{} develops the corresponding nonlinear model.
First, we smooth the initial Gaussian field on a scale $R$ and declare a collapsed region wherever the smoothed value exceeds $\delta_c$.
Adding a long mode $\delta_L$ lowers the threshold seen by the Gaussian field;
the fraction of collapsed regions can be written as the tail integral
\begin{equation}
 F(\delta_L)=\frac{1}{\sqrt{2\pi}\sigma}
 \int_{\delta_c-\delta_L}^{\infty}\dd\delta_s\,
 e^{-\delta_s^2/(2\sigma^2)}
 =\frac12\operatorname{erfc}\!\left(
 \frac{\delta_c-\delta_L}{\sqrt2\sigma}\right),
 \qquad \bar F\equiv F(0),
 \qquad
 \nu_c\equiv\frac{\delta_c}{\sigma(R)},
 \label{eq:threshold-tail}
\end{equation}
where $\sigma^2(R)$ is the field variance.
Differentiating the lower limit gives the Lagrangian response,
\begin{equation}
 b_1^{\rm L}
 \equiv\left.\frac{1}{\bar F}\frac{\partial F}{\partial\delta_L}\right|_{\delta_L=0}
 =\frac{e^{-\nu_c^2/2}}{\sqrt{2\pi}\,\sigma(R)\,\bar F}.
 \label{eq:thresholdbias}
\end{equation}
\cref{fig:thresholdbias} illustrates the threshold shift and the resulting abundance response.
Rare high-threshold objects therefore respond most strongly.
Number conservation under the Lagrangian-to-Eulerian map~\cite{MoWhite} gives
\begin{equation}
 1+\delta_g^{\rm E}=(1+\delta_m)(1+\delta_g^{\rm L}).
 \label{eq:eulerian-bias-conversion}
\end{equation}
At linear order $\delta_g^{\rm E}=\delta_m+\delta_g^{\rm L}$, and hence
\begin{equation}
 b_1^{\rm E}=1+b_1^{\rm L},
 \qquad
 \delta_g(\vx)=b_1\delta_m(\vx)+\epsilon(\vx).
 \label{eq:linearbiasepsilon}
\end{equation}
From here on $b_1$ always means the Eulerian bias.
The stochastic field $\epsilon$ carries the small-scale formation information that the long-wavelength matter field does not determine, and at leading order $\langle\epsilon\delta_m\rangle=0$ with power that is approximately constant.
Ideal Poisson sampling gives $P_\epsilon=1/\bar n_g$, although real tracers need not be Poisson~\cite{BaldaufStoch}.

\begin{warningbox}{$\delta_c$ and halo morphology}
In \cref{eq:threshold-tail,eq:thresholdbias}, $\delta_c$ is the threshold used to illustrate abundance bias;
it does not imply that selected regions remain spherical during collapse.
An ellipsoidal-collapse barrier~\cite{ShethMoTormen} changes the predicted abundance and bias coefficients, but not the response mechanism:
a long-wavelength density perturbation shifts the effective threshold and changes the number of selected objects.
\end{warningbox}

\begin{keybox}{Lecture 2---from collapse to observation}
Spherical collapse provides a nonlinear clock, the Zel'dovich deformation tensor explains the origin of the cosmic web, and shell crossing marks the limit of the pressureless single-stream description.
Selecting collapsed environments gives $\delta_g=b_1\delta_m+\epsilon$.
\lectureref{3} derives the effect of radial velocities on this galaxy field and its compression into power spectrum multipoles.
\end{keybox}

\phantomsection
\label{exercises:lecture2}
\subsection*{Exercises}
\begin{enumerate}
\item Evaluate \cref{eq:deltalsphere} at turnaround, $\eta=\pi$, and collapse, $\eta=2\pi$.
Explain why the nonlinear density diverges while the extrapolated linear density remains finite.
\item Use \cref{eq:psilinear,eq:za-deformation} to expand \cref{eq:zadensity} to first order and obtain $\delta_{\rm ZA}=\delta^{(1)}+\Order(D_+^2)$.
\item For $\nu_c\gg1$, differentiate \cref{eq:threshold-tail} to derive \cref{eq:thresholdbias}, then use $\bar F\simeq e^{-\nu_c^2/2}/(\sqrt{2\pi}\,\nu_c)$ to obtain the leading response.
How does this result interpret the bias of rare objects?
\item Working consistently to first order in $\delta_i$, start from \cref{eq:growing-mode-initialization} and derive the $5/3$ coefficient in \cref{eq:spherical-energy}.
Then combine $A=-GM/(2E)$ with $A^3=GMB^2$ to show $t_{\rm coll}=2\pi B\propto\delta_i^{-3/2}$ and explain why the extrapolated $\delta_c$ is independent of $\delta_i$.
\end{enumerate}

\lecturetitle{3}{The galaxy power spectrum multipoles and Gaussian covariance}

\begin{objectives}
Starting from \cref{eq:linearbiasepsilon}, students will
\begin{itemize}
\item derive the Jacobian from real to redshift space and obtain the Kaiser field and power spectrum;
\item derive the first three even Legendre polynomials from orthogonality;
\item derive $P_0$, $P_2$ and $P_4$;
\item construct a shell estimator and obtain its Gaussian covariance from Wick's theorem.
\end{itemize}
Shot noise and mode-count conventions will remain explicit throughout.
\end{objectives}

\subsection{Real-space galaxies and stochastic power}

At leading order and on scales well above the formation scale of the tracer, we adopt the linear bias expansion~\cite{BiasReview}:
\begin{equation}
 \delta_g(\vk)=b_1\delta_m(\vk)+\epsilon(\vk),
 \qquad
 \langle\epsilon(\vk)\delta_m(\vk')\rangle=0,
 \label{eq:linearbiask}
\end{equation}
where the second condition imposes no correlation between $\epsilon$ and the long-wavelength matter field.
Squaring and averaging gives the real-space galaxy power spectrum.
Under ideal Poisson sampling, the leading stochastic power is $1/\bar n_g$~\cite{BiasReview}:
\begin{equation}
 P_{gg}(k)=b_1^2\Plin(k)+P_\epsilon(k),
 \qquad P_\epsilon(k)\simeq P_N\simeq\frac{1}{\bar n_g}.
 \label{eq:realgalaxypower}
\end{equation}
The clustering signal must remain distinct from the total variance of a measured mode.
A shot-noise-subtracted estimator has mean $b_1^2\Plin$, whereas its variance still depends on $b_1^2\Plin+P_N$:
subtracting a known mean does not remove the random fluctuation.

\subsection{Distances inferred from redshift}

A survey infers radial distance from the observed redshift rather than measuring it directly.
We work in the plane-parallel or distant-observer approximation:
a single fixed line of sight $\nhat$ applies to the full survey, and peculiar velocity displaces only the radial coordinate.
Ref.~\cite{Hamilton1998} provides the classic treatment of linear redshift-space distortions and their multipoles.
With $\vu=\dd\vx/\dd\tau$, the redshift-space coordinate is
\begin{equation}
 \vs=\vx+\frac{u_\parallel(\vx)}{\cH}\nhat,
 \qquad u_\parallel\equiv\vu\cdot\nhat.
 \label{eq:rsdmap}
\end{equation}
The map interprets the Doppler contribution to redshift as Hubble expansion and thereby displaces the inferred radial coordinate.
It does not change the physical galaxy positions at the time of observation.

Counting the same galaxies in either coordinate system gives number conservation,
\begin{equation}
 [1+\delta_g^s(\vs)]\dd^3s=[1+\delta_g(\vx)]\dd^3x.
 \label{eq:numberconservationrsd}
\end{equation}
Expanding the Jacobian of \cref{eq:rsdmap} to linear order in the velocity gives
\begin{equation}
 \det\!\left(\frac{\partial s_i}{\partial x_j}\right)
 =1+\frac{1}{\cH}\partial_\parallel u_\parallel,
 \qquad
 \delta_g^s=\delta_g-\frac{1}{\cH}\partial_\parallel u_\parallel.
 \label{eq:rsdjac}
\end{equation}

\subsection{Continuity supplies the Kaiser enhancement}

For the growing irrotational mode, \cref{eq:thetaf} fixes the velocity in terms of the density alone,
\begin{equation}
 \vu(\vk)=\ii\cH f\frac{\vk}{k^2}\delta_m(\vk).
 \label{eq:velocityfourier}
\end{equation}
Define the cosine between the wavevector and the line of sight by
\begin{equation}
 \mu\equiv\khat\cdot\nhat.
 \label{eq:mudef}
\end{equation}
The two Fourier signs that control the result must be tracked explicitly.
With the Fourier convention of \cref{eq:fourier},
\begin{equation}
 \partial_\parallel\longrightarrow \ii k\mu,
 \qquad
 u_\parallel(\vk)=\ii\cH f\frac{\mu}{k}\delta_m(\vk),
 \qquad
 -\frac{1}{\cH}\partial_\parallel u_\parallel
 =-\frac{(\ii k\mu)(\ii\cH f\mu/k)}{\cH}\delta_m
 =+f\mu^2\delta_m.
 \label{eq:kaiser-sign-chain}
\end{equation}
The explicit minus sign from the Jacobian cancels the product $\ii^2=-1$.
Thus the Fourier transform of \cref{eq:rsdjac} becomes
\begin{equation}
 \delta_g^s(\vk)=\left(b_1+f\mu^2\right)\delta_m(\vk)+\epsilon(\vk).
 \label{eq:kaiserfield}
\end{equation}
Coherent infall produces the plus sign:
compression along the inferred radial coordinate increases the inferred overdensity.
Squaring the Kaiser factor gives the linear signal;
adding the stochastic contribution gives the total power in a measured mode:
\begin{align}
 P_g^s(k,\mu)&=\left(b_1+f\mu^2\right)^2\Plin(k),
 \label{eq:kaiserpower}\\
 P_{\rm tot}(k,\mu)&=P_g^s(k,\mu)+P_N.
 \label{eq:totalpower}
\end{align}
The Kaiser power spectrum is anisotropic, but it is symmetric under $\mu\to-\mu$, and only even angular multipoles can therefore appear.

\begin{warningbox}{fingers of God beyond the Kaiser limit}
\cref{eq:kaiserpower} retains only the coherent infall of \cref{eq:velocityfourier}, for which every galaxy in a patch shares one velocity.
Galaxies in a virialized halo instead have orbital velocities with a dispersion of several hundred $\mathrm{km\,s^{-1}}$ that is uncorrelated with the long mode.
Interpreting these random velocities as distances extends each halo into a radial streak---the finger-of-God effect---and suppresses power at large $\mu$ and high $k$.
This suppression opposes the coherent Kaiser enhancement and can make the observed quadrupole change sign at quasi-nonlinear $k$.
The range of validity of \cref{eq:kaiserpower} depends on the tracer;
velocity dispersion becomes important at larger scales for denser and more massive halos.
We do not introduce a phenomenological damping factor.
Because the long-wavelength theory cannot predict this small-scale dispersion, the \extensionref{} represents its effect through the counterterms and stochastic terms of \cref{eq:rsdcounterterms,eq:galaxystoch}.
\end{warningbox}

\subsection{Legendre decomposition}

The Legendre polynomials form an orthogonal basis for the anisotropy on $-1\leq\mu\leq1$~\cite{Hamilton1998}:
\begin{equation}
 \int_{-1}^{1}\dd\mu\,\Lpoly_\ell(\mu)\Lpoly_{\ell'}(\mu)
 =\frac{2}{2\ell+1}\delta^K_{\ell\ell'},
 \label{eq:legendreorth}
\end{equation}
whose first even members are
\begin{equation}
 \Lpoly_0=1,
 \qquad
 \Lpoly_2=\frac12(3\mu^2-1),
 \qquad
 \Lpoly_4=\frac18(35\mu^4-30\mu^2+3).
 \label{eq:legendrefirst}
\end{equation}
We may therefore expand any sufficiently regular axisymmetric power spectrum in this basis and recover it from its coefficients,
\begin{equation}
 P(k,\mu)=\sum_{\ell=0}^{\infty}P_\ell(k)\Lpoly_\ell(\mu),
 \qquad
 P_\ell(k)=\frac{2\ell+1}{2}\int_{-1}^{1}\dd\mu\,
 \Lpoly_\ell(\mu)P(k,\mu).
 \label{eq:multipoleprojection}
\end{equation}
The monopole is an orientation average, the quadrupole measures the leading line-of-sight contrast, and the hexadecapole measures the next even angular pattern.
The calculation in \cref{sec:linear-multipoles} also uses the inverse relations, which follow directly from \cref{eq:legendrefirst}:
\begin{equation}
 \mu^2=\frac13\Lpoly_0+\frac23\Lpoly_2,
 \qquad
 \mu^4=\frac15\Lpoly_0+\frac47\Lpoly_2+\frac{8}{35}\Lpoly_4.
 \label{eq:mupowerlegendre}
\end{equation}

\subsection{The linear monopole, quadrupole, and hexadecapole}
\label{sec:linear-multipoles}

First expand the Kaiser power spectrum of \cref{eq:kaiserpower} in powers of $\mu$:
\begin{equation}
 P_g^s=(b_1^2+2b_1f\mu^2+f^2\mu^4)\Plin.
 \label{eq:kaiserexpanded}
\end{equation}
Substituting the inversions of \cref{eq:mupowerlegendre} gives the three surviving multipoles,
\begin{align}
 P_0(k)&=\left(b_1^2+\frac23b_1f+\frac15f^2\right)\Plin(k),
 \label{eq:P0kaiser}\\
 P_2(k)&=\left(\frac43b_1f+\frac47f^2\right)\Plin(k),
 \label{eq:P2kaiser}\\
 P_4(k)&=\frac{8}{35}f^2\Plin(k).
 \label{eq:P4kaiser}
\end{align}
No higher multipoles appear at this order because the angular dependence stops at $\mu^4$, and isotropic shot noise adds $P_N$ to the total monopole alone.

These three expressions cannot separate every parameter.
Each term contains the product of a bias or growth factor with $\Plin$, and $\Plin\propto\sigma_8^2$ by \cref{eq:sigma8}, so the multipoles depend on $b_1$, $f$ and $\sigma_8$ only through the two combinations $b_1\sigma_8$ and $f\sigma_8$.
No amount of linear data breaks that degeneracy:
doubling $\sigma_8$ while halving both $b_1$ and $f$ leaves \crefrange{eq:P0kaiser}{eq:P4kaiser} unchanged.
A redshift survey therefore reports $f\sigma_8$ rather than $f$.
The one-loop terms introduced in the \extensionref{} scale differently with $\sigma_8$ and begin to break this degeneracy.

\begin{derivationbox}{three immediate checks}
Setting $f=0$ leaves only the real-space monopole $b_1^2\Plin$, as it must, while setting $b_1=0$ instead leaves velocity selection alone and still produces $P_0=f^2\Plin/5$, $P_2=4f^2\Plin/7$ and $P_4=8f^2\Plin/35$, with $P_4$ carrying no $b_1$ at all at linear order.
These are algebraic checks on the projection, not claims that either tracer is realistic.
\end{derivationbox}

\subsection{A shell estimator and the number of modes}

For a periodic volume, define modes normalized by the box volume as
\begin{equation}
 \delta_V(\vq)\equiv V^{-1/2}\int_V\dd^3x\,
 e^{-\ii\vq\cdot\vx}\delta(\vx),
 \label{eq:box-field}
\end{equation}
so a statistically homogeneous box field obeys the discrete two-point rule
\begin{equation}
 \big\langle\delta_V(\vq)\delta_V(\vq')\big\rangle
 =\delta^K_{\vq,-\vq'}P_{\rm tot}(\vq).
 \label{eq:box-two-point}
\end{equation}
The factor $V^{-1/2}$ makes the mode variance a power spectrum rather than a power spectrum times a volume.
We let bin $i$ cover $k_i^-<q<k_i^+$.
The exact full-shell count and its continuum and thin-shell approximations are
\begin{equation}
 N_i^{\rm full}
 \equiv\sum_{\substack{\vq\\ k_i^-<q<k_i^+}}1
 \simeq\frac{V}{(2\pi)^3}\frac{4\pi}{3}\left[(k_i^+)^3-(k_i^-)^3\right]
 \simeq\frac{Vk_i^2\Delta k_i}{2\pi^2},
 \label{eq:Nfull}
\end{equation}
where $\Delta k_i\equiv k_i^+-k_i^-$ in the final approximation.
The count includes both $\vq$ and $-\vq$.
Subtracting shot noise, we define the multipole estimator in a periodic box~\cite{FKP,Yamamoto} as
\begin{equation}
 \widehat P_\ell(k_i)=\frac{2\ell+1}{N_i^{\rm full}}
 \sum_{\vq\in i}\Lpoly_\ell(\mu_{\vq})
 \left[|\delta_{g,V}^s(\vq)|^2-P_N\right].
 \label{eq:multipoleestimator}
\end{equation}
Its continuum angular average returns \cref{eq:multipoleprojection}.
Subtracting $P_N$ changes the estimator mean, but it does not change the mode variance $P_{\rm tot}=P_g^s+P_N$ in \cref{eq:box-two-point}.
Because $\delta(-\vq)=\delta^*(\vq)$, however, the number of statistically independent complex modes is only
\begin{equation}
 N_i^{\rm ind}=\frac12N_i^{\rm full}.
 \label{eq:Nind}
\end{equation}
The full-shell and independent-half-shell counts are equivalent when paired with their respective prefactors.
Mixing one prefactor with the other mode count produces the factor-of-two error described in \cref{sec:covariance-blocks}.

\subsection{Why the covariance is a four-point function}

Each power estimate is quadratic in $\delta$~\cite{Huterer}, so the covariance of two estimates contains four fields.
For a zero-mean Gaussian field, Wick's theorem factorizes the four-point function into three products of two-point functions:
\begin{equation}
 \langle1234\rangle_G
 =\langle12\rangle\langle34\rangle
 +\langle13\rangle\langle24\rangle
 +\langle14\rangle\langle23\rangle.
 \label{eq:wick}
\end{equation}
\begin{derivationbox}{from the estimator to the covariance prefactor}
We write $a_\ell=2\ell+1$ and substitute \cref{eq:multipoleestimator} twice.
Constants such as the subtracted $P_N$ have no covariance, so the connected estimator product is
\[
 \frac{a_\ell a_{\ell'}}{N_i^{\rm full}N_j^{\rm full}}
 \sum_{\vq\in i}\sum_{\vq'\in j}
 \Lpoly_\ell(\mu_{\vq})\Lpoly_{\ell'}(\mu_{\vq'})
 \left[\langle\delta_{\vq}\delta_{-\vq}
 \delta_{\vq'}\delta_{-\vq'}\rangle
 -P_{\rm tot}(\vq)P_{\rm tot}(\vq')\right].
\]
The first Wick pairing in \cref{eq:wick} is exactly the subtracted product of means.
Using \cref{eq:box-two-point}, the other two pairings impose respectively $\vq'=-\vq$ and $\vq'=+\vq$.
They are equal for the even multipoles considered here, and they can contribute only when $i=j$.
Hence
\[
 \Cov^G_{\ell\ell'}
 =\delta^K_{ij}\frac{2a_\ell a_{\ell'}}{(N_i^{\rm full})^2}
 \sum_{\vq\in i}\Lpoly_\ell(\mu_{\vq})
 \Lpoly_{\ell'}(\mu_{\vq})P_{\rm tot}^2(\vq).
\]
Finally, a full shell has a uniform angular distribution, so $(N_i^{\rm full})^{-1}\sum_{\vq\in i}h(\mu_{\vq})\to \tfrac12\int_{-1}^{1}\dd\mu\,h(\mu)$.
The factor $2$ from the two surviving Wick pairings cancels this $1/2$, leaving the prefactor of the master equation below.
\end{derivationbox}
The Gaussian covariance is
\begin{equation}
 \boxed{
 \Cov^G_{\ell\ell'}(k_i,k_j)
 =\delta^K_{ij}\,
 \frac{(2\ell+1)(2\ell'+1)}{N_i^{\rm full}}
 \int_{-1}^{1}\dd\mu\,
 \Lpoly_\ell(\mu)\Lpoly_{\ell'}(\mu)
 [P_g^s(k_i,\mu)+P_N]^2.}
 \label{eq:covmaster}
\end{equation}
Different multipoles are correlated because they reweight the same finite set of anisotropic modes.
Setting $P_g^s=P$, independent of $\mu$, and $\ell=\ell'=0$, gives the consistency check
\begin{equation}
 \operatorname{Var}\widehat P_0=\frac{2(P+P_N)^2}{N_i^{\rm full}}.
 \label{eq:isotropicvariance}
\end{equation}

\subsection{All six covariance blocks for \texorpdfstring{$\ell=0,2,4$}{ell=0,2,4}}
\label{sec:covariance-blocks}

We fold the isotropic noise into the total monopole,
\begin{equation}
 S_0=P_0+P_N,
 \qquad S_2=P_2,
 \qquad S_4=P_4,
 \label{eq:Sdefs}
\end{equation}
and define the angular integral
\begin{equation}
 Q_{\ell\ell'}\equiv
 \frac{(2\ell+1)(2\ell'+1)}{2}
 \int_{-1}^{1}\dd\mu\,
 \Lpoly_\ell\Lpoly_{\ell'}
 \left[\sum_{L=0,2,4}S_L\Lpoly_L(\mu)\right]^2,
 \label{eq:Qdef}
\end{equation}
in terms of which the Gaussian covariance takes the compact form
\begin{equation}
 \Cov^G_{\ell\ell'}(k_i,k_j)
 =\delta^K_{ij}\frac{2}{N_i^{\rm full}}Q_{\ell\ell'}(k_i).
 \label{eq:covQ}
\end{equation}

\begin{warningbox}{equivalent mode-count conventions}
By \cref{eq:hermitian}, $\vq$ and $-\vq$ form one Hermitian pair.
A full shell counts both entries, whereas an independent half-shell counts one, so \cref{eq:Nind} gives
\[
 N_i^{\rm ind}=\frac12N_i^{\rm full}.
\]
For the even multipoles considered here, the same Gaussian covariance can therefore be written as
\[
 \frac{2Q_{\ell\ell'}}{N_i^{\rm full}}
 =\frac{Q_{\ell\ell'}}{N_i^{\rm ind}}.
\]
We use the full-shell convention throughout.
Combining $2Q_{\ell\ell'}$ with $N_i^{\rm ind}$ would mix the two conventions and double the covariance.
\cref{eq:isotropicvariance} explicitly checks this normalization.
\end{warningbox}

The required integrals over four Legendre polynomials are evaluated in \cref{app:covariance-coefficients}.
Substituting them into \cref{eq:Qdef} gives the six independent blocks~\cite{NguyenAkitsuTaruya},
\begin{align}
Q_{00}={}&S_0^2+\frac15S_2^2+\frac19S_4^2,
\label{eq:Q00}\\
Q_{02}={}&2S_0S_2+\frac27S_2^2+\frac47S_2S_4+\frac{100}{693}S_4^2,
\label{eq:Q02}\\
Q_{04}={}&2S_0S_4+\frac{18}{35}S_2^2+\frac{40}{77}S_2S_4
+\frac{162}{1001}S_4^2,
\label{eq:Q04}\\
Q_{22}={}&5S_0^2+\frac{20}{7}S_0S_2+\frac{20}{7}S_0S_4
+\frac{15}{7}S_2^2+\frac{120}{77}S_2S_4+\frac{8945}{9009}S_4^2,
\label{eq:Q22}\\
Q_{24}={}&\frac{36}{7}S_0S_2+\frac{200}{77}S_0S_4
+\frac{108}{77}S_2^2+\frac{3578}{1001}S_2S_4+\frac{900}{1001}S_4^2,
\label{eq:Q24}\\
Q_{44}={}&9S_0^2+\frac{360}{77}S_0S_2+\frac{2916}{1001}S_0S_4
+\frac{16101}{5005}S_2^2+\frac{3240}{1001}S_2S_4
+\frac{42849}{17017}S_4^2.
\label{eq:Q44}
\end{align}
The symmetry of \cref{eq:Qdef} under $\ell\leftrightarrow\ell'$ gives $Q_{20}=Q_{02}$, $Q_{40}=Q_{04}$ and $Q_{42}=Q_{24}$.
Ref.~\cite{NguyenAkitsuTaruya} writes the same six blocks as a bilinear form in two spectra, which is what a multi-tracer covariance needs;
\crefrange{eq:Q00}{eq:Q44} are its single-tracer case.

\subsection{From a periodic box to a finite survey}

A real survey has a finite and anisotropic window, spatial weights, and a line of sight that turns with position, so translation invariance is broken and neighbouring $k$ bins couple~\cite{Wilson}.
In the organization of Ref.~\cite{WadekarScoccimarro}, the continuous Gaussian term factorizes cosmology from survey geometry as
\begin{equation}
 C^{G({\rm cont})}_{\ell_1\ell_2}(k_1,k_2)
 =\sum_{L_1,L_2}P_{L_1}(k_1)P_{L_2}(k_2)
 W^{(1)}_{\ell_1,\ell_2,L_1,L_2}(k_1,k_2).
 \label{eq:windowcov}
\end{equation}
Discrete tracers add mixed power--noise kernels $W^{(2)}$ and pure-noise kernels $W^{(3)}$.
In a periodic box these terms reduce to the single shift $S_0=P_0+P_N$;
in a survey their weights differ and must remain separate.
Packages such as \texttt{thecov} implement this generalization for a finite survey window~\cite{ForeroSanchez}.

\subsection{From multipoles to likelihoods and Fisher forecasts}

For $N_k$ wavenumber bins, define the bin-major data vector
\begin{equation}
 \widehat{\bm P}\equiv
 \bigl(\widehat P_0(k_1),\widehat P_2(k_1),\widehat P_4(k_1),
 \ldots,\widehat P_0(k_{N_k}),\widehat P_2(k_{N_k}),
 \widehat P_4(k_{N_k})\bigr)^{\mathsf T}.
 \label{eq:multipole-data-vector}
\end{equation}
The model vector $\bm P(\bm\theta)$ uses the same ordering for parameters $\bm\theta$.
Define $N_{\rm d}=3N_k$ and the residual $\Delta\bm P\equiv\widehat{\bm P}-\bm P(\bm\theta)$.
For a periodic box, assembling \cref{eq:covQ} gives
\begin{equation}
 C_{(i\ell)(j\ell')}(\bm\theta)
 =\delta^K_{ij}\frac{2Q_{\ell\ell'}(k_i;\bm\theta)}{N_i^{\rm full}},
 \qquad \ell,\ell'\in\{0,2,4\}.
 \label{eq:multipole-covariance-matrix}
\end{equation}
A finite survey replaces this block-diagonal matrix with the window-coupled covariance of \cref{eq:windowcov}.

For parameter-dependent mean and covariance, the normalized Gaussian likelihood is~\cite{Huterer}
\begin{equation}
 -2\ln\mathcal L(\bm\theta)
 =\Delta\bm P^{\mathsf T}\mathsf C^{-1}\Delta\bm P
 +\ln\det\mathsf C+N_{\rm d}\ln(2\pi),
 \label{eq:gausslike}
\end{equation}
where $\mathsf C$ denotes the matrix in \cref{eq:multipole-covariance-matrix} or its survey generalization.
The determinant term is required when $\mathsf C$ varies with $\bm\theta$.
Define $\bm P_{,\alpha}\equiv\partial\bm P/\partial\theta_\alpha$ and $\mathsf C_{,\alpha}\equiv\partial\mathsf C/\partial\theta_\alpha$, evaluated at fiducial parameters $\bm\theta_{\rm fid}$.
Differentiating \cref{eq:gausslike} and averaging over realizations gives the Fisher matrix
\begin{equation}
 \begin{split}
 F_{\alpha\beta}
 &\equiv-\left\langle
 \frac{\partial^2\ln\mathcal L}
 {\partial\theta_\alpha\partial\theta_\beta}\right\rangle\\
 &=\bm P_{,\alpha}^{\mathsf T}\mathsf C^{-1}\bm P_{,\beta}
 +\frac12\operatorname{Tr}\!\left[
 \mathsf C^{-1}\mathsf C_{,\alpha}
 \mathsf C^{-1}\mathsf C_{,\beta}\right].
 \end{split}
 \label{eq:fisher-matrix}
\end{equation}
Fixing the covariance at the fiducial parameters sets $\mathsf C(\bm\theta)\to\mathsf C(\bm\theta_{\rm fid})$ and $\mathsf C_{,\alpha}=0$.
The determinant is then constant, and \cref{eq:fisher-matrix} reduces to
\[
 F_{\alpha\beta}^{\rm fixed\ C}
 =\bm P_{,\alpha}^{\mathsf T}\mathsf C^{-1}\bm P_{,\beta}.
\]
The marginalized Cram\'er--Rao bound is $\sigma_{\rm marg}(\theta_\alpha)\geq\sqrt{(F^{-1})_{\alpha\alpha}}$.

If $\mathsf C$ is estimated from $N_{\rm mock}$ mock catalogues, the Hartlap rescaling~\cite{Hartlap}
\[
 \mathsf C^{-1}\longrightarrow
 \frac{N_{\rm mock}-N_{\rm d}-2}{N_{\rm mock}-1}\mathsf C^{-1}
\]
removes the mean bias of its inverse but does not propagate covariance-estimation uncertainty.
The periodic-box Gaussian covariance is a baseline~\cite{Grieb}.
For a finite survey, window coupling, the connected trispectrum, and super-sample covariance~\cite{TakadaHu} can modify $\mathsf C$ without changing the assembly of the likelihood.

\begin{figure}[t]
\centering
\includegraphics[width=0.88\textwidth]{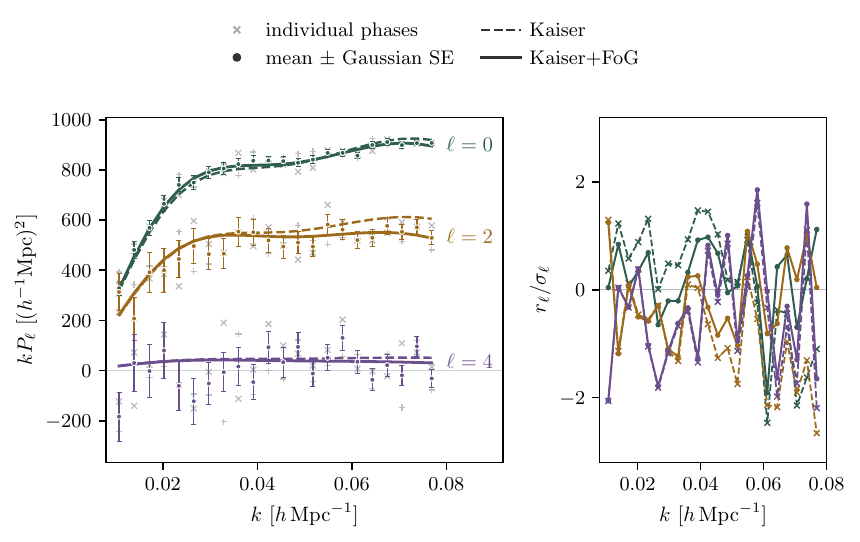}
\caption{Linear redshift-space models fitted to AbacusPNG halo multipoles.
\emph{Left:} $kP_\ell$ for $\ell=0,2,4$ from two $2\,h^{-1}\mathrm{Gpc}$ phases at $z=0.5$;
gray crosses show the individual phases and colored points show their mean with Gaussian standard errors.
\emph{Right:} diagonal residuals $r_\ell/\sigma_\ell$, where $r_\ell=P_\ell^{\rm mean}-P_\ell^{\rm model}$.
Both models are fitted jointly over $0.01\leq k\leq0.08\,h\,\mathrm{Mpc}^{-1}$ at fixed cosmology, with $f=0.759$.
They fit $b_1$ and the monopole stochastic amplitude;
the Kaiser+FoG model also multiplies the Kaiser power by $D_{\rm FoG}(k,\mu)=[1+(k\mu\sigma_v)^2/2]^{-1}$.
The pure Kaiser fit gives $b_1=1.206$ and $\chi^2/N_{\rm dof}=1.44$.
Adding $\sigma_v=5.95\,h^{-1}\mathrm{Mpc}$ gives $b_1=1.235$ and $\chi^2/N_{\rm dof}=1.16$.
The monopole is the raw halo monopole, with the fitted constant stochastic term included in both curves.}
\label{fig:kaisermultipoles}
\end{figure}

\cref{fig:kaisermultipoles} tests the Kaiser model against halo multipoles from two $f_{\rm NL}=0$ phases of the AbacusPNG simulations~\cite{AbacusPNG}.
Each phase has a periodic box of side $2\,h^{-1}\mathrm{Gpc}$ at $z=0.5$.
We select halos with $M\geq10^{12}\,h^{-1}M_\odot$.
The fits hold the simulation cosmology fixed and use $0.01\leq k\leq0.08\,h\,\mathrm{Mpc}^{-1}$.
They fit the raw halo monopole together with the quadrupole and hexadecapole, so the constant stochastic contribution is inferred rather than subtracted.
For the equally weighted phase mean, the covariance is $\mathsf C_{\rm mean}=(\mathsf C_{\rm ph000}+\mathsf C_{\rm ph001})/4$.
The error bars in \cref{fig:kaisermultipoles} are Gaussian standard errors from its diagonal, while the fits use the full $3\times3$ multipole block in each $k$ bin.
The stored covariance counts independent half-space modes;
the half-shell relation in \cref{eq:Nind} makes this convention equivalent to the full-shell count used in \cref{eq:covQ}.
The limited accuracy of the linear model motivates the \extensionref{}.

\begin{keybox}{from the core lectures to the advanced extension}
Lectures 1--3 give the complete linear model:
\[
 \Plin(k)\longrightarrow(b_1,f,P_N)\longrightarrow
 P_g^s(k,\mu)\longrightarrow(P_0,P_2,P_4)\longrightarrow\Cov^G.
\]
It remains valid only while nonlinear mode coupling, scale-dependent bias, nonlinear velocities and BAO smearing are negligible.
The \extensionref{} incorporates these effects while isolating uncontrolled small-scale physics from large-scale inference.
\end{keybox}

\phantomsection
\label{exercises:lecture3}
\subsection*{Exercises}
\begin{enumerate}
\item Apply Gram--Schmidt to $1,\mu^2,\mu^4$ with inner product $\langle f,g\rangle=\int_{-1}^{1}\dd\mu\,f(\mu)g(\mu)$ and impose $\Lpoly_\ell(1)=1$.
Then directly project the Kaiser power to derive \crefrange{eq:P0kaiser}{eq:P4kaiser} without using \cref{eq:mupowerlegendre}.
\item Set $S_2=S_4=0$ in \cref{eq:Qdef} and derive $Q_{\ell\ell'}=(2\ell+1)S_0^2\delta^K_{\ell\ell'}$.
Explain both the nonzero diagonal variances and the zero off-diagonal blocks.
\item For a zero-mean Gaussian field in a narrow isotropic shell, write one non-self-conjugate mode as $\delta_{g,V}^s=X+\ii Y$, with independent $X,Y$ of variance $P_{\rm tot}/2$ (constant across the shell).
Derive $\operatorname{Var}|\delta_{g,V}^s|^2=P_{\rm tot}^2$, then derive the monopole variances using an independent half-shell and a full shell.
Why does subtracting known $P_N$ change the mean but not the variance?
\end{enumerate}

\extensiontitle{Nonlinear bias and one-loop EFT power spectrum}

\begin{objectives}
Readers of this extension should be able to
\begin{itemize}
\item identify why the linear model fails;
\item organize nonlinear evolution with Eulerian perturbation theory;
\item derive the $P_{22}$ and $P_{13}$ loop structure;
\item explain why a long-wavelength theory requires counterterms;
\item build the renormalized galaxy-bias basis through cubic order;
\item assemble the one-loop redshift-space galaxy power spectrum before its angular projection.
\end{itemize}
\end{objectives}

\subsection{What must change beyond the Kaiser model}

The linear approximation evolves each Fourier mode independently.
Nonlinear gravity does not:
the terms dropped from \cref{eq:continuityfull,eq:eulerfull} couple wavevectors together.
On quasi-linear scales the coupling shows up in four distinct ways:
\begin{enumerate}
\item broadband mode coupling changes the shape of the matter power spectrum;
\item galaxy formation responds nonlinearly and tidally to the matter environment;
\item the velocity mapping generates angular powers beyond $\mu^4$;
\item long displacements smear the BAO feature.
\end{enumerate}
Small-scale multi-streaming lies outside the pressureless fluid description, motivating the effective field theory of large-scale structure (EFTofLSS).
Ref.~\cite{IvanovEFT} provides a focused pedagogical treatment of its renormalization and infrared resummation, while Refs.~\cite{BaumannNicolis,EFT} contain the foundational effective-fluid construction.
The EFT retains calculable long modes and parameterizes the influence of unresolved short modes through all operators allowed by the symmetries.

\subsection{Eulerian perturbation theory}

Returning to the order expansion of \cref{eq:fluid-perturbative-series}, define the normalized velocity divergence~\cite{Bernardeau,IvanovGGI},
\begin{equation}
 \delta=\sum_{n\geq1}\delta^{(n)},
 \qquad
 \Theta\equiv-\frac{\theta}{\cH f}=\sum_{n\geq1}\Theta^{(n)},
 \qquad \delta^{(1)}=\Theta^{(1)},
 \label{eq:sptseries}
\end{equation}
where the kernels $F_n$ and $G_n$ collect the mode couplings at each order.
Under the EdS-kernel approximation, $F_n$ and $G_n$ retain their EdS forms, while $D_+$ and $f$ vary with cosmology.
\begin{align}
 \delta^{(n)}(\vk)&=
 \int\prod_{a=1}^{n}\frac{\dd^3k_a}{(2\pi)^3}
 (2\pi)^3\dirac\!\left(\vk-\sum_a\vk_a\right)
 F_n(\vk_1,\ldots,\vk_n)\prod_a\delta^{(1)}(\vk_a),
 \label{eq:Fndef}\\
 \Theta^{(n)}(\vk)&=
 \int\prod_{a=1}^{n}\frac{\dd^3k_a}{(2\pi)^3}
 (2\pi)^3\dirac\!\left(\vk-\sum_a\vk_a\right)
 G_n(\vk_1,\ldots,\vk_n)\prod_a\delta^{(1)}(\vk_a).
 \label{eq:Gndef}
\end{align}
The cosmological time dependence is then carried by $D_+^n$ and powers of $f$.
Define $\mu_{12}=\hat\vk_1\cdot\hat\vk_2$ as the cosine between two incoming modes.
The symmetrized second-order coefficients follow directly from the fluid equations.

\begin{derivationbox}{second-order fluid recursion}
In EdS, set $y\equiv\ln a$, so $\partial_\tau=\cH\partial_y$, and use $\Theta=-\theta/\cH$.
Define the common convolution measure by $\int_{12}^{\vk}\equiv\int\dd^3k_1/(2\pi)^3\,\dd^3k_2/(2\pi)^3 (2\pi)^3\dirac(\vk-\vk_1-\vk_2)$.
Fourier transforming the nonlinear continuity and Euler equations and using Poisson gives~\cite{Bernardeau,IvanovGGI}
\begin{equation}
 \begin{aligned}
 \partial_y\delta(\vk)-\Theta(\vk)
 &=\int_{12}^{\vk}\alpha(\vk_1,\vk_2)\Theta(\vk_1)\delta(\vk_2),\\
 \partial_y\Theta(\vk)+\frac12\Theta(\vk)-\frac32\delta(\vk)
 &=\int_{12}^{\vk}\beta(\vk_1,\vk_2)\Theta(\vk_1)\Theta(\vk_2).
 \end{aligned}
 \label{eq:spt-mode-coupling}
\end{equation}
Here $\alpha(\vk_1,\vk_2)\equiv(\vk_1+\vk_2)\cdot\vk_1/k_1^2$ and $\beta(\vk_1,\vk_2)\equiv |\vk_1+\vk_2|^2(\vk_1\cdot\vk_2)/(2k_1^2k_2^2)$.
At second order, the growing solution scales as $a^2$.
Inserting \cref{eq:Fndef,eq:Gndef} into \cref{eq:spt-mode-coupling} and defining $\alpha_s\equiv[\alpha(\vk_1,\vk_2)+\alpha(\vk_2,\vk_1)]/2$ gives
\begin{equation}
 2F_2-G_2=\alpha_s,
 \qquad \frac52G_2-\frac32F_2=\beta.
 \label{eq:spt-second-order-system}
\end{equation}
Solving gives $F_2=5\alpha_s/7+2\beta/7$ and $G_2=3\alpha_s/7+4\beta/7$.
With $r\equiv k_1/k_2$, the vertices become $\alpha_s=1+\mu_{12}(r+r^{-1})/2$ and $\beta=\mu_{12}(r+r^{-1})/2+\mu_{12}^2$;
substituting them gives the kernels below.
\end{derivationbox}

The symmetrized second-order kernels are
\begin{align}
 F_2(\vk_1,\vk_2)&=\frac57
 +\frac12\mu_{12}\left(\frac{k_1}{k_2}+\frac{k_2}{k_1}\right)
 +\frac27\mu_{12}^2,
 \label{eq:F2}\\
 G_2(\vk_1,\vk_2)&=\frac37
 +\frac12\mu_{12}\left(\frac{k_1}{k_2}+\frac{k_2}{k_1}\right)
 +\frac47\mu_{12}^2.
 \label{eq:G2}
\end{align}
The constant pieces encode local growth, the ratio term is advection, and the $\mu_{12}^2$ piece is tidal~\cite{Goroff,JainBertschinger}.
The difference between $F_2$ and $G_2$ shows that density and velocity cease to be interchangeable beyond first order.

\subsection{Wick contractions and one-loop matter power spectrum}

Inserting \cref{eq:sptseries} into $\langle\delta\delta\rangle$ and retaining terms through fourth order in the linear field gives
\begin{equation}
 P_{mm}^{\rm SPT}(k)=\Plin(k)+P_{22}(k)+P_{13}(k)+\Order(\Plin^3),
 \label{eq:oneloopspt}
\end{equation}
where
\begin{align}
 P_{22}(k)&=2\intq F_2^2(\vq,\vk-\vq)
 \Plin(q)\Plin(|\vk-\vq|),
 \label{eq:P22}\\
 P_{13}(k)&=6\Plin(k)\intq F_3(\vk,\vq,-\vq)\Plin(q).
 \label{eq:P13}
\end{align}
The numerical factors count Wick contractions:
$P_{22}$ correlates two second-order fields, while $P_{13}$ correlates a first-order field with a third-order one.
Each contribution contains one unconstrained loop momentum $\vq$, defining the one-loop order~\cite{MakinoSasakiSuto,ScoccimarroFrieman}.

\begin{derivationbox}{why there are only two one-loop shapes}
Expanding both external fields and retaining four powers of the linear density gives
\begin{equation}
 \left.\langle\delta\delta\rangle\right|_{(\delta^{(1)})^4}
 =\langle\delta^{(2)}\delta^{(2)}\rangle
 +2\langle\delta^{(1)}\delta^{(3)}\rangle.
 \label{eq:oneloop-order-four}
\end{equation}
The third-order correlator $\langle\delta^{(1)}\delta^{(2)}\rangle$ vanishes because an odd Gaussian moment is zero.
In the $22$ term either linear field on the left can pair with the first field on the right, after which the last pairing is fixed:
two Wick contractions give the coefficient $2$.
In the $13$ term there are two choices for which external field is cubic and three choices for which of its linear factors pairs with the lone field, giving $2\times3=6$.
Momentum conservation fixes the remaining pair to $(\vq,-\vq)$, but leaves the magnitude and direction of $\vq$ free.
That single surviving momentum integral is the loop in \cref{eq:P22,eq:P13}.
\end{derivationbox}

\subsection{Infrared cancellation and ultraviolet sensitivity}

For $q\ll k$, the individual loop contributions contain large displacement terms.
An equal-time correlator cannot depend on a uniform displacement, however, and the equivalence principle therefore forces the leading infrared pieces of $P_{22}$ and $P_{13}$ to cancel exactly~\cite{JainBertschinger,PelosoPietroni}.
Finite long-wavelength displacements nevertheless remain important for the BAO.
For a smooth spectrum, the soft expansion is controlled by $q/k$.
For the oscillatory part $P_{\rm w}(k)\sim\sin(k r_{\rm BAO})$, each $k$ derivative brings down $r_{\rm BAO}$, so the expansion is instead controlled by $q r_{\rm BAO}=q/k_{\rm osc}$, where $k_{\rm osc}\equiv r_{\rm BAO}^{-1}$.
Modes with $k_{\rm osc}\lesssim q\ll k$ are therefore soft relative to $k$ but can shift the BAO phase by order unity.
At the other limit, $q\gg k$, the integrals sample short modes where neither the single-stream fluid nor finite-order perturbation theory is reliable.
Consequently, even a formally finite integral contains uncontrolled small-scale contributions.
A cutoff makes this ultraviolet sensitivity explicit, but physical predictions cannot depend on the cutoff value.

\subsection{Coarse-graining and the EFT counterterm}

\begin{figure}[t]
\centering
\includegraphics[width=0.80\textwidth]{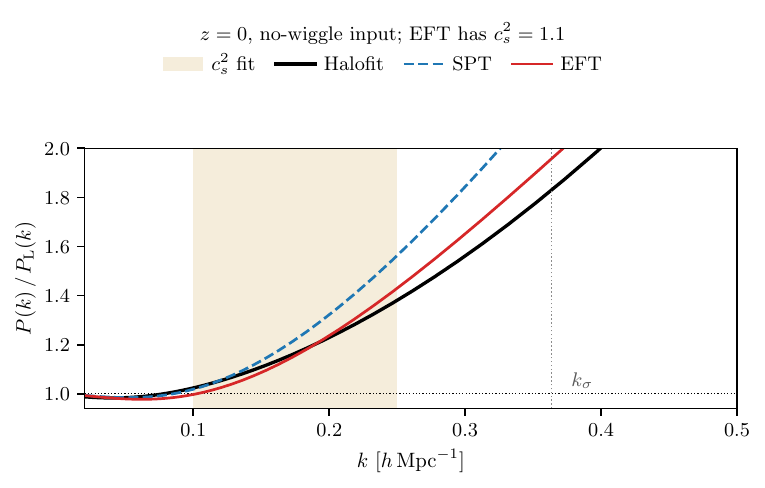}
\caption{Effect of the leading EFT counterterm.
All three curves are divided by $\Plin$ at $z=0$, with a no-wiggle input so that the broadband is not obscured by acoustic oscillations.
One-loop SPT overshoots the Halofit target and continues to rise;
adding $-2c_s^2k^2\Plin$ with a single $c_s^2\simeq1.1\,(h^{-1}\mathrm{Mpc})^2$ brings the prediction into agreement.
The coefficient $c_s^2$ is fitted to Halofit over the shaded $0.10$--$0.25\,h\,\mathrm{Mpc}^{-1}$ interval, so the close EFT agreement demonstrates the role of the counterterm but is not an independent accuracy test.
The dotted line denotes Halofit's nonlinear scale, $k_\sigma=0.364\,h\,\mathrm{Mpc}^{-1}$.
This scale marks the loss of finite-order perturbative control, not a validity boundary for the fitting formula.}
\label{fig:eftvsspt}
\end{figure}

\cref{fig:eftvsspt} illustrates the leading counterterm correction.
To derive its origin, smooth the equations at a scale $\Lambda^{-1}$ larger than the nonlinear length.
Smoothing a nonlinear product differs from multiplying the smoothed fields.
This difference generates an effective stress tensor in the long-wavelength Euler equation.
Symmetry allows us to expand that stress in derivatives, whose leading deterministic scalar correction is proportional to $\nabla^2\delta$.
The leading field-level term is~\cite{CarrascoForeman}
\begin{equation}
 \delta_{\rm ct}(\vk)=-c_s^2k^2\delta^{(1)}(\vk)
 \quad\Longrightarrow\quad
 2\langle\delta^{(1)}\delta_{\rm ct}\rangle
 =-2c_s^2k^2\Plin(k).
 \label{eq:counterterm-field}
\end{equation}
The factor $2$ counts the pair of cross-correlations.
The matter power spectrum therefore becomes
\begin{equation}
 P_{mm}^{\rm EFT}(k)
 =\Plin+P_{22}+P_{13}-2c_s^2k^2\Plin+P_{\rm stoch}^{m}+\cdots.
 \label{eq:mattereft}
\end{equation}
The coefficient $c_s^2$ is a renormalized parameter with dimensions of length squared.
It absorbs the leading response of unresolved small-scale structure.
Despite the name, it is not the microscopic sound speed of cold dark matter.
A change in smoothing cutoff shifts the loop contribution and $c_s^2$ separately, but their sum remains unchanged at the order retained.
Mass and momentum conservation suppress the large-scale matter stochastic contribution~\cite{MercolliPajer} as $P_{\rm stoch}^{m}\propto k^4$.
A discrete tracer obeys no analogous conservation constraint, so its stochastic power may begin with a constant.

\subsection{A renormalized nonlinear galaxy-bias expansion}

\begin{figure}[t]
\centering
\includegraphics[width=0.90\textwidth]{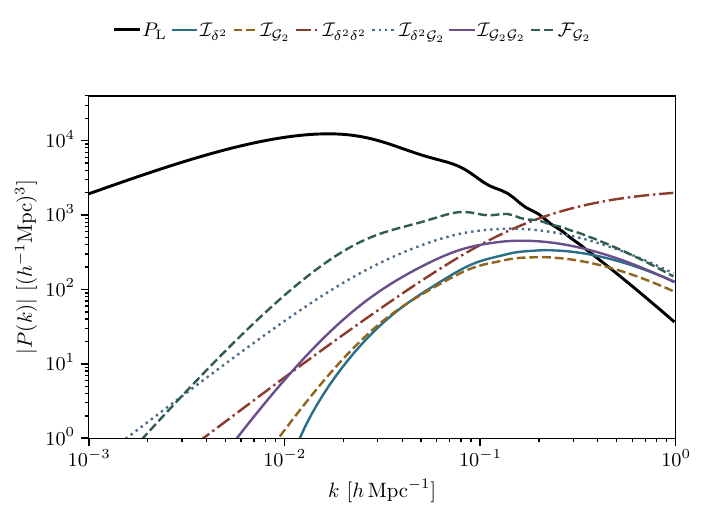}
\caption{One-loop contributions from nonlinear bias operators.
The integrals follow from the operators of \cref{eq:nonlinearbias} at $z=0.61$, with infrared resummation applied.
The vertical axis shows absolute magnitudes in power-spectrum units, and $\Plin$ provides a scale reference.
Each subscript names the contraction:
$\mathcal{I}_{\delta^2}$ and $\mathcal{I}_{\Gtwo}$ cross $[\delta^2]$ and $[\Gtwo]$ with the linear density, the double subscripts give the auto- and cross-spectra of the two quadratic operators, and $\mathcal{F}_{\Gtwo}$ is the $P_{13}$-type term in which $[\Gtwo]$ closes on a loop against $\delta$.
Absolute values are plotted, since several of these integrals are negative, and no bias coefficients multiply them.
The curves therefore isolate the $k$ dependence supplied by each operator, while $b_2$, $b_{\Gtwo}$ and their products determine its contribution to $P_{gg}$.
The operator-dependent curves become comparable to $\Plin$ across the quasi-nonlinear regime and dominate by $k$ of order $1\,h\,\mathrm{Mpc}^{-1}$ in this illustrative calculation.
The steep low-$k$ behaviour of $\mathcal{I}_{\delta^2\delta^2}$ appears after the $k\to0$ constant of the bare convolution $\int\Plin\Plin$ is subtracted and absorbed into the renormalized stochastic amplitude $P_{\epsilon}^{\{0\}}$.
By contrast, \cref{eq:renormdelta2} displays the mean and lower-order operator subtractions.}
\label{fig:biasbreakdown}
\end{figure}

Galaxy formation is local only in an effective, long-wavelength sense.
The equivalence principle forbids any dependence on a uniform potential or a uniform acceleration, leaving the tidal tensor as the first locally observable gravitational field, $\partial_i\partial_j\Phi$.
Through cubic order and leading derivatives, we use the Eulerian operator basis~\cite{BiasReview}
\begin{equation}
 \begin{split}
 \delta_g={}&b_1\delta
 +\frac{b_2}{2}[\delta^2]
 +b_{\Gtwo}[\Gtwo]
 +\frac{b_3}{6}[\delta^3]
 +b_{\delta\Gtwo}[\delta\Gtwo]
 +b_{\Gthree}[\Gthree]\\
 &+b_{\nabla^2\delta}R_*^2\nabla^2\delta
 +\epsilon+\epsilon_\delta\delta+\cdots.
 \end{split}
 \label{eq:nonlinearbias}
\end{equation}
We define the tidal operator by~\cite{ChanScoccimarroSheth}
\begin{equation}
 \Gtwo(\Phi_g)=
 (\partial_i\partial_j\Phi_g)^2-(\nabla^2\Phi_g)^2,
 \qquad \nabla^2\Phi_g=\delta.
 \label{eq:G2operator}
\end{equation}
Other papers use the trace-free tidal invariant $K_{ij}K_{ij}$, and because the two bases differ by a multiple of $\delta^2$, the numerical value called $b_2$ changes with the basis.
We define the cubic operator $\Gthree$ as the difference between the same invariant built from the density and velocity potentials~\cite{BiasReview}:
\begin{equation}
 \Gthree=\Gtwo(\Phi_g)-\Gtwo(\Phi_v),
 \qquad \nabla^2\Phi_v=-\frac{\theta}{f\cH},
 \label{eq:G3operator}
\end{equation}
with $\theta$ the velocity divergence.
The two potentials agree at linear order, so $\Gthree$ starts at third order and measures how far the velocity has decoupled from the density.
The opposite ordering of the two terms is also in common use, so the sign of a fitted $b_{\Gthree}$ must be interpreted together with the convention adopted by its source, as for $b_2$ above.
The square brackets in \cref{eq:nonlinearbias} denote renormalized composite operators rather than bare products.
Operationally, we define them by
\begin{equation}
 [O]\equiv O-\sum_{O'\,\text{lower order}}Z_{OO'}(\Lambda)O'.
 \label{eq:renormalized-brackets}
\end{equation}
The coefficients $Z_{OO'}$ are chosen so that every cutoff-dependent long-mode piece proportional to a lower-order operator $O'$ is subtracted.
The operator $\delta^2$, for example, has a nonzero mean, and its correlation with a long mode contains a cutoff-dependent piece proportional to $\sigma^2(\Lambda)\Plin(k)$.
Retaining its mean and leading mixing with $\delta$ gives~\cite{McDonald2006,Assassi}
\begin{equation}
 [\delta^2]=\delta^2-\langle\delta^2\rangle
 -\frac{68}{21}\sigma^2(\Lambda)\delta+\cdots.
 \label{eq:renormdelta2}
\end{equation}
The subtraction moves the cutoff-dependent contributions into $b_1$, thereby defining $b_1$ as the physical large-scale response.
Renormalization is necessary to define the coefficients of the bias expansion.
\cref{fig:biasbreakdown} displays the contribution supplied by each operator in \cref{eq:nonlinearbias} after these subtractions.

The derivative term in \cref{eq:nonlinearbias} records the finite formation scale $R_*$, while the stochastic fields record whatever the long-wavelength operators fail to determine.
For later use after the redshift-space map, we parameterize the stochastic galaxy power through $\Order(k^2)$ as~\cite{BiasReview,DAmico}
\begin{equation}
 P_{\rm stoch}^g(k,\mu)=P_{\epsilon}^{\{0\}}
 +k^2P_{\epsilon}^{\{2\}}+k^2\mu^2P_{\epsilon,\mu}^{\{2\}}+\cdots.
 \label{eq:galaxystoch}
\end{equation}

\subsection{The exact redshift-space map and the \texorpdfstring{$Z_n$}{Zn} kernels}

Beyond the Kaiser limit, we retain the plane-parallel map of \cref{eq:rsdmap} without linearizing its velocity dependence.
Fourier transforming number conservation and changing variables from $\vs$ to $\vx$ gives
\begin{equation}
 \begin{split}
 (2\pi)^3\dirac(\vk)+\delta_g^s(\vk)
 & =\int\dd^3x\,e^{-\ii\vk\cdot\vs(\vx)}[1+\delta_g(\vx)]\\
 & =\int\dd^3x\,e^{-\ii\vk\cdot\vx}[1+\delta_g(\vx)]
 e^{-\ii k_\parallel u_\parallel(\vx)/\cH}.
 \end{split}
 \label{eq:rsd-fourier-number-conservation}
\end{equation}
Number conservation has already cancelled the coordinate Jacobian in the measure, and no expansion in $\delta_g$ or $u_\parallel$ has been made.
Subtracting the zero mode $\int\dd^3x\,e^{-\ii\vk\cdot\vx}=(2\pi)^3\dirac(\vk)$ gives the exact Fourier-space field,
\begin{equation}
 \delta_g^s(\vk)=\int\dd^3x\,e^{-\ii\vk\cdot\vx}
 \left\{[1+\delta_g(\vx)]
 e^{-\ii k_\parallel u_\parallel(\vx)/\cH}-1\right\}.
 \label{eq:rsdexact}
\end{equation}
Expanding the density, velocity and exponential together organizes the result into a series of redshift-space kernels:
\begin{equation}
 \delta_g^s(\vk)=\sum_{n\geq1}
 \int\prod_{a=1}^{n}\frac{\dd^3k_a}{(2\pi)^3}
 (2\pi)^3\dirac\!\left(\vk-\sum_a\vk_a\right)
 Z_n(\vk_1,\ldots,\vk_n)\prod_a\delta^{(1)}(\vk_a).
 \label{eq:Zndef}
\end{equation}
The lowest kernel is the Kaiser factor derived in \lectureref{3}~\cite{Kaiser,ScoccimarroCouchmanFrieman}:
\begin{equation}
 Z_1(\vk)=b_1+f\mu^2.
 \label{eq:Z1}
\end{equation}
For $\vk=\vk_1+\vk_2$, define $\mu_i=\hat\vk_i\cdot\nhat$ and $\mu=\hat\vk\cdot\nhat$.
Using the bias normalization of \cref{eq:nonlinearbias}~\cite{ClassPT}, the quadratic redshift-space kernel is
\begin{equation}
 \begin{split}
 Z_2(\vk_1,\vk_2)={}&b_1F_2+f\mu^2G_2+\frac{b_2}{2}
 +b_{\Gtwo}\left(\mu_{12}^2-1\right)\\
 &+\frac{f\mu k}{2}
 \left[\frac{\mu_1}{k_1}(b_1+f\mu_2^2)
 +\frac{\mu_2}{k_2}(b_1+f\mu_1^2)\right].
 \end{split}
 \label{eq:Z2}
\end{equation}
\paragraph{Origins of the $Z_2$ terms.}
The four groups of terms have distinct physical origins:
\begin{center}
\small
\begin{tabularx}{0.96\linewidth}{@{}l X@{}}
\toprule
Term & Origin \\ \midrule
$b_1F_2$ & second-order density and gravitational evolution \\
$f\mu^2G_2$ & second-order velocity divergence in the radial Jacobian \\
$b_2/2$ and $b_{\Gtwo}(\mu_{12}^2-1)$ & local quadratic and tidal galaxy response \\
the final $f\mu k/2$ bracket & expansion of the coordinate-mapping exponential in
\cref{eq:rsdexact} \\
\bottomrule
\end{tabularx}
\end{center}
The tidal angular factor follows from the operator definition.
In Fourier space,
\[
 \partial_i\partial_j\Phi(\vk)=-k_i k_j\Phi(\vk),
 \qquad \nabla^2\Phi_g=\delta,
\]
so the kernel of $(\partial_i\partial_j\Phi_g)^2-(\nabla^2\Phi_g)^2$ is
\[
 \frac{(\vk_1\cdot\vk_2)^2}{k_1^2k_2^2}-1=\mu_{12}^2-1.
\]
A full $Z_3$ expression is unnecessary for this term-by-term explanation.
It contains analogous cubic gravity, velocity, mapping and bias pieces but introduces no new category of term~\cite{IvanovSimonovicZaldarriaga};
the one-loop power spectrum uses it only in the special contraction $Z_3(\vk,\vq,-\vq)$.

\subsection{The assembled one-loop redshift-space power spectrum}

The raw one-loop signal for a single tracer is
\begin{equation}
 \begin{split}
 P_{gg}^{s,\rm raw}(k,\mu)={}&Z_1^2(\vk)\Plin(k)
 +2\intq Z_2^2(\vq,\vk-\vq)\Plin(q)\Plin(|\vk-\vq|)\\
 &+6Z_1(\vk)\Plin(k)\intq Z_3(\vk,\vq,-\vq)\Plin(q).
 \end{split}
 \label{eq:oneloopredshift}
\end{equation}
The renormalized EFT model adds effective counterterms and stochastic terms.
Adopting the field-level factorized basis of Ref.~\cite{DAmico} and absorbing basis-dependent factors into fitted coefficients gives
\begin{equation}
 P_{\rm ctr}^s(k,\mu)
 =-2k^2\Plin(k)\left(c_0+c_2\mu^2+c_4\mu^4+c_6\mu^6\right).
 \label{eq:rsdcounterterms}
\end{equation}
The four coefficients are constrained because the redshift-space map multiplies three field-level angular coefficients by the Kaiser factor.
Setting $x=\mu^2$, this factorization reads
\begin{equation}
 c_0+c_2x+c_4x^2+c_6x^3
 =(b_1+fx)(r_0+r_2x+r_4x^2).
 \label{eq:counterterm-polynomial}
\end{equation}
Matching powers gives $c_0=b_1r_0$, $c_2=b_1r_2+fr_0$, $c_4=b_1r_4+fr_2$, and $c_6=fr_4$.
Successive elimination yields
\[
 r_0=\frac{c_0}{b_1},\qquad
 r_2=\frac{b_1c_2-fc_0}{b_1^2},\qquad
 r_4=\frac{b_1^2c_4-b_1fc_2+f^2c_0}{b_1^3}.
\]
Eliminating the three $r_i$ coefficients fixes
\begin{equation}
 c_6=\frac{f}{b_1^3}\left(c_4b_1^2-c_2b_1f+c_0f^2\right).
 \label{eq:c6fixed}
\end{equation}
Only three $k^2$ counterterm coefficients are independent;
adding a fourth coefficient would discard the field-level factorization that produced them.
After projection onto retained $\ell=0,2,4$, the direct basis of Refs.~\cite{ClassPT,ChudaykinIvanov}, $-2k^2\Plin(\tilde c_0+\tilde c_2f\mu^2+\tilde c_4f^2\mu^4)$, is equivalent to the factorized basis.
The two bases are not equal as full functions of $\mu$ because the factorized basis generally contains an $\ell=6$ component.
Whether powers of $f$ are absorbed into the coefficients is a basis choice.
The figure below uses this latter basis, whose counterterm polynomial stops at $\mu^4$.
Higher-derivative galaxy bias is degenerate with directions in this counterterm space rather than, in general, with the monopole coefficient alone.
The counterterm sector also encodes the finger-of-God effects described in \lectureref{3}.
The $\mu^4$ and $\mu^6$ terms of \cref{eq:rsdcounterterms}, together with the $k^2\mu^2$ stochastic term of \cref{eq:galaxystoch}, encode the leading effect of small-scale velocity dispersion on the long-wavelength power spectrum.
Their fitted coefficients replace a chosen damping function and treat the dispersion as unresolved short-distance physics.

\begin{figure}[t]
\centering
\includegraphics[width=0.96\textwidth]{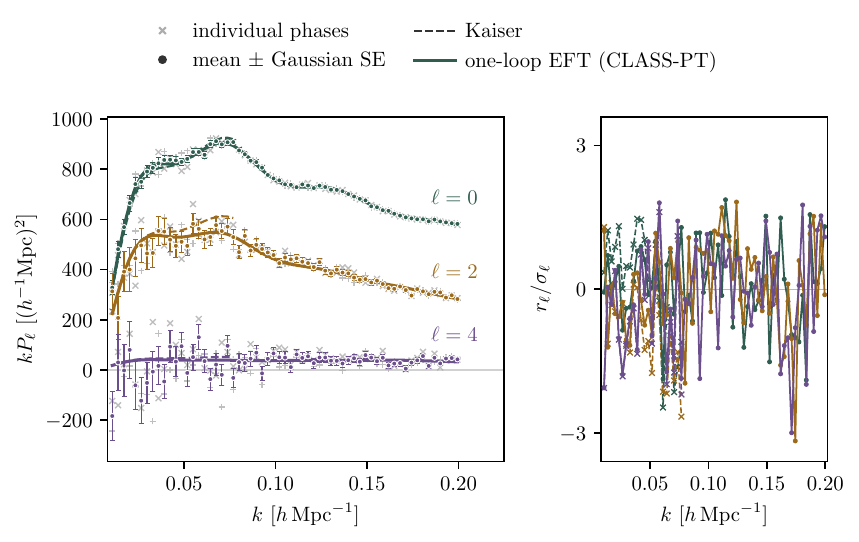}
\caption{Linear Kaiser and one-loop EFT fits to AbacusPNG halo multipoles.
\emph{Left:} the two phases and their mean are shown as in \cref{fig:kaisermultipoles}.
The dashed Kaiser curve ends at its fit cut, $k_{\max}=0.08\,h\,\mathrm{Mpc}^{-1}$, whereas the solid curve is the supplied CLASS-PT one-loop EFT best fit, including the infrared resummation of \cref{sec:ir-resummation}, through $k_{\max}=0.20\,h\,\mathrm{Mpc}^{-1}$~\cite{ClassPT}.
\emph{Right:} diagonal residuals relative to the same phase-mean Gaussian standard errors.
The cosmology is fixed to the simulation value in both fits.
The joint one-loop fit to the two phases gives $b_1=1.254$ and $\chi^2/N_{\rm dof}=1.012$ for 357 degrees of freedom.
The curves are best-fit predictions;
the error bars describe the data covariance and do not include parameter or perturbative uncertainty.}
\label{fig:rsdmultipoles}
\end{figure}

Collecting signal, counterterms and stochastic contributions, the complete fixed-order power spectrum is
\begin{equation}
 P_{gg}^{s,\rm EFT}=P_{gg}^{s,\rm raw}+P_{\rm ctr}^s+P_{\rm stoch}^g.
 \label{eq:fullfixedeft}
\end{equation}
The loops generate even powers of $\mu$ through $\mu^8$ before counterterms and projection are applied, so the mean contains multipoles beyond $\ell=4$ even if an analysis retains only $\ell=0,2,4$.
\cref{fig:rsdmultipoles} compares the assembled one-loop model with the same AbacusPNG multipoles used in \cref{fig:kaisermultipoles}.
The linear Kaiser curve stops at its fitted $k_{\max}=0.08\,h\,\mathrm{Mpc}^{-1}$, while the CLASS-PT one-loop EFT fit uses $k_{\max}=0.20\,h\,\mathrm{Mpc}^{-1}$.

\subsection{Infrared resummation of the BAO}
\label{sec:ir-resummation}

Long displacements blur the oscillatory BAO component while weakly affecting the smooth broadband power spectrum.
We therefore split the linear matter power spectrum as
\begin{equation}
 \Plin=P_{\rm nw}+P_{\rm w}.
 \label{eq:wigglesplit}
\end{equation}
Following Ref.~\cite{IvanovSibiryakov}, let $k_S\simeq0.2\,h\,\mathrm{Mpc}^{-1}$ separate the resummed long modes from the short modes, and let $j_\ell$ denote a spherical Bessel function.
Define $k_{\rm osc}\equiv r_{\rm BAO}^{-1}$ with $r_{\rm BAO}\simeq110\,h^{-1}\mathrm{Mpc}$, so $q/k_{\rm osc}=q\,r_{\rm BAO}$:
\begin{align}
 \Sigma^2&=\frac{1}{6\pi^2}\int_0^{k_S}\dd q\,P_{\rm nw}(q)
 \left[1-j_0\!\left(\frac{q}{k_{\rm osc}}\right)
 +2j_2\!\left(\frac{q}{k_{\rm osc}}\right)\right],
 \label{eq:Sigma}\\
 \delta\Sigma^2&=\frac{1}{2\pi^2}\int_0^{k_S}\dd q\,P_{\rm nw}(q)
 j_2\!\left(\frac{q}{k_{\rm osc}}\right),
 \label{eq:dSigma}\\
 \Sigma_{\rm tot}^2(\mu)&=
 [1+f\mu^2(2+f)]\Sigma^2
 +f^2\mu^2(\mu^2-1)\delta\Sigma^2.
 \label{eq:Sigmatot}
\end{align}
Let $P_{1\text{-loop}}^s[P]$ denote the two loop integrals in \cref{eq:oneloopredshift} evaluated with $P$ in place of $\Plin$.
At next-to-leading order, the resummed power spectrum can be written as~\cite{IvanovSibiryakov}
\begin{equation}
 \begin{split}
 P_{gg}^{s,\rm IR}={}&Z_1^2\left[P_{\rm nw}
 +e^{-k^2\Sigma_{\rm tot}^2}(1+k^2\Sigma_{\rm tot}^2)P_{\rm w}\right]\\
 &+P_{1\text{-loop}}^s[P_{\rm nw}]
 +e^{-k^2\Sigma_{\rm tot}^2}
 \left(P_{1\text{-loop}}^s[\Plin]-P_{1\text{-loop}}^s[P_{\rm nw}]\right)\\
 &+P_{\rm ctr}^s+P_{\rm stoch}^g.
 \end{split}
 \label{eq:IRassembled}
\end{equation}
The factor $1+k^2\Sigma_{\rm tot}^2$ prevents double counting of the one-loop displacement.

\begin{derivationbox}{fixed-order check of the IR organization}
Define $X\equiv k^2\Sigma_{\rm tot}^2=\Order(\Plin)$.
Expanding the compensated tree factor and defining $\Delta P_{1\text{-loop}}^s\equiv P_{1\text{-loop}}^s[\Plin]-P_{1\text{-loop}}^s[P_{\rm nw}]$ gives
\begin{equation}
 \begin{aligned}
 e^{-X}(1+X)&=1-\frac{X^2}{2}+\cdots=1+\Order(\Plin^2),\\
 P_{1\text{-loop}}^s[P_{\rm nw}]+e^{-X}\Delta P_{1\text{-loop}}^s
 &=P_{1\text{-loop}}^s[\Plin]+\Order(\Plin^3).
 \end{aligned}
 \label{eq:ir-fixed-order-check}
\end{equation}
The first line differs from unity only at $\Order(\Plin^2)$ and multiplies $P_{\rm w}=\Order(\Plin)$;
the second uses $\Delta P_{1\text{-loop}}^s=\Order(\Plin^2)$.
Both corrections therefore begin beyond one loop.
\end{derivationbox}

The expansion in \cref{eq:ir-fixed-order-check} recovers the fixed-order result \cref{eq:fullfixedeft}.
Full cumulant derivations and alternative organizations are given in Refs.~\cite{SenatoreZaldarriaga,Blas2016}.

\subsection{Power counting and the trust scale}

\begin{figure}[t]
\centering
\includegraphics[width=0.88\textwidth]{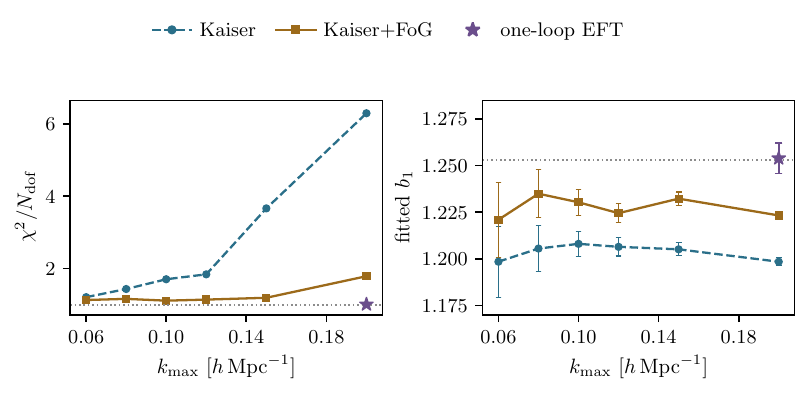}
\caption{Fit quality and fitted bias as functions of the scale cut.
\emph{Left:} reduced $\chi^2$ after refitting the Kaiser and Kaiser+FoG models at each $k_{\max}$;
the star shows the CLASS-PT one-loop EFT fit at $0.20\,h\,\mathrm{Mpc}^{-1}$, and the dotted line marks unity.
\emph{Right:} fitted $b_1$ with conditional $1\sigma$ errors at fixed cosmology;
the horizontal line marks the supplied reference $b_1=1.2531$.
The Kaiser fit deteriorates as higher-$k$ bins are added;
at $0.20\,h\,\mathrm{Mpc}^{-1}$, Kaiser+FoG gives $\chi^2/N_{\rm dof}=1.79$, compared with $1.012$ for the one-loop fit.
These values describe this two-phase halo sample and do not define a universal $k_{\max}$.}
\label{fig:multipoles}
\end{figure}

Additional parameters do not justify fitting arbitrarily small scales.
A controlled EFT analysis specifies its nonlinear scale, perturbative order, and estimate of the first omitted term.
We choose $k_{\max}$ where that estimate remains below the required accuracy, then test fit quality and parameter stability as the cut varies.
Fit stability is necessary, but it does not replace an estimate of the first omitted term.
If the required accuracy follows the statistical error, increasing the survey volume moves the cut to lower $k$ at fixed perturbative order.
\cref{fig:multipoles} illustrates this distinction for the same two AbacusPNG phases.

\begin{keybox}{physical model before numerical evaluation}
A complete galaxy model requires renormalized nonlinear bias, higher derivatives, stochasticity, redshift-space kernels, counterterms and IR resummation in addition to $P_{22}+P_{13}$.
Once these ingredients are specified, only efficient evaluation of the loop integrals remains.
\end{keybox}

\subsection{From the physical model to numerical multipoles}

\cref{eq:oneloopredshift} contains a three-dimensional loop convolution for every cosmology, redshift, external wavenumber, angular value and bias sector.
Repeating direct quadrature at every likelihood evaluation is therefore too expensive.
FFTLog expands the input spectrum in complex power laws, after which the loop convolutions become precomputable matrix contractions~\cite{Hamilton2000,FFTLog}.
This change of numerical basis leaves the EFT model and the split $\Plin=P_{\rm nw}+P_{\rm w}$ unchanged.

The two internal spectra in $P_{22}$ produce a double contraction $\sum_{mn}c_mc_nM_{22}^{mn}$.
The single internal spectrum in $P_{13}$ instead produces $\Plin(k)\sum_m c_mM_{13}^m$.
Appendix~\ref{app:master-integrals} derives the scalar and line-of-sight master integrals behind these contractions and states the minimal numerical checks.

Perturbation theory predicts the full anisotropic power spectrum $P_g^s(k,\mu)$ rather than an individual multipole.
\cref{fig:pipeline} summarizes the implementation order:
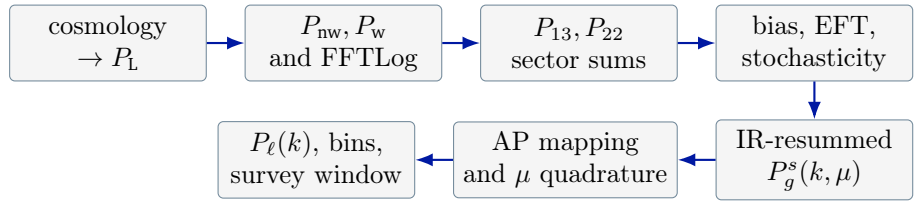
\begin{figure}[htbp]
\centering
\begin{cmfigurefont}
\begin{tikzpicture}[>=Latex,node distance=5mm and 5mm,
 every node/.style={font=\small}]
\tikzset{pipe/.style={draw=Ink!55,rounded corners=2pt,fill=SoftGray,
 minimum height=10mm,minimum width=26mm,align=center}}
\node[pipe] (cos) {cosmology\\$\to\Plin$};
\node[pipe,right=of cos] (split) {$P_{\rm nw},P_{\rm w}$\\and FFTLog};
\node[pipe,right=of split] (loops) {$P_{13},P_{22}$\\sector sums};
\node[pipe,right=of loops] (model) {bias, EFT,\\stochasticity};
\node[pipe,below=of model] (ir) {IR-resummed\\$P_g^s(k,\mu)$};
\node[pipe,left=of ir] (ap) {AP mapping\\and $\mu$ quadrature};
\node[pipe,left=of ap] (pell) {$P_\ell(k)$, bins,\\survey window};
\draw[->,thick,LinkBlue] (cos)--(split);
\draw[->,thick,LinkBlue] (split)--(loops);
\draw[->,thick,LinkBlue] (loops)--(model);
\draw[->,thick,LinkBlue] (model)--(ir);
\draw[->,thick,LinkBlue] (ir)--(ap);
\draw[->,thick,LinkBlue] (ap)--(pell);
\end{tikzpicture}
\end{cmfigurefont}
\caption{Calculation sequence for the one-loop model.
Appendix~\ref{app:master-integrals} accelerates the loop evaluation;
multipole projection follows assembly of the anisotropic physical model.}
\label{fig:pipeline}
\end{figure}
The final projection onto Legendre multipoles is
\begin{equation}
 P_\ell(k)=\frac{2\ell+1}{2}\int_{-1}^{1}\dd\mu\,
 \Lpoly_\ell(\mu)P_g^{s,\rm IR}(k,\mu).
 \label{eq:advancedmultipoles}
\end{equation}
When required, Alcock--Paczynski scaling~\cite{AlcockPaczynski} changes the true $(k,\mu)$ entering the model and adds a volume Jacobian before projection.
Finite $k$ bins and the survey window are then applied to the theoretical multipoles so that the prediction matches the measurement operator.

\begin{keybox}{the complete physical and numerical chain}
\[
 \begin{aligned}
 \mathcal R&\to\Plin\to\{\text{collapse, web, }b_1\}
 \to P_g^s(k,\mu)\to\{P_0,P_2,P_4,\Cov\}\\
 &\to P_g^{s,\rm one\mbox{-}loop\ EFT}
 \xrightarrow{\rm FFTLog}P_\ell^{\rm fast}(k).
 \end{aligned}
\]
Lectures 1--3 give the leading-order chain.
The advanced extension specifies the nonlinear bias and one-loop EFT model, while Appendix~\ref{app:master-integrals} evaluates its loop contributions and states the minimal numerical checks.
\end{keybox}

\phantomsection
\label{exercises:extension}
\subsection*{Exercises}
\begin{enumerate}
\item Set $b_1=1$ and all nonlinear biases to zero.
Expand \cref{eq:rsdexact} through second order and derive only the coordinate-mapping bracket of \cref{eq:Z2}.
Check explicitly that the result is symmetric under $\vk_1\leftrightarrow\vk_2$.
\item Show why the large-scale stochastic matter power has no white-noise term, whereas discrete tracers may retain one.
For a localized matter rearrangement $\Delta\rho(\vx)$, Fourier transform and expand its low-$k$ moments.
Use mass and momentum conservation to derive $\delta_{\rm stoch}^m=\Order(k^2)$ and $P_{\rm stoch}^m=\Order(k^4)$.
Why do the same constraints not apply to tracer number?
\end{enumerate}

\section*{\color{scipostdeepblue}{Acknowledgements}}
I thank the staff of the International Centre for Interdisciplinary Science and Education (ICISE) for their hospitality and Rencontres du Vietnam for its support of the 10th Vietnam School of Astrophysics (VSOA10).
I also thank the students and fellow lecturers for making the school an enjoyable and memorable experience.

\paragraph{Funding information}
NMN acknowledges support from the JSPS KAKENHI Grant Numbers JP25K23373 and JP26H00404.
This work was supported by the World Premier International Research Center Initiative (WPI), MEXT, Japan.

\clearpage
\phantomsection
\addcontentsline{toc}{section}{Appendices}
\markboth{Appendices}{Appendices}
\begin{center}
{\LARGE\bfseries\color{Ink} Appendices}\par
\vspace{5pt}\color{Ink!45}\rule{0.92\textwidth}{0.5pt}
\end{center}
\appendix

\section{Generating and checking the periodic-box covariance}
\label{app:covariance-coefficients}

For $\ell,\ell',L_1,L_2\in\{0,2,4\}$ define the exact rational tensor
\begin{equation}
 A_{\ell\ell';L_1L_2}
 \equiv\frac{(2\ell+1)(2\ell'+1)}{2}
 \int_{-1}^{1}\dd\mu\,
 \Lpoly_\ell\Lpoly_{\ell'}\Lpoly_{L_1}\Lpoly_{L_2}.
 \label{eq:Atensor}
\end{equation}
Then
\begin{equation}
 Q_{\ell\ell'}=
 \sum_{L_1,L_2\in\{0,2,4\}}
 A_{\ell\ell';L_1L_2}S_{L_1}S_{L_2}.
 \label{eq:Qgenerator}
\end{equation}
For implementation, the small rational tensor can be generated once from explicit polynomials, checked against its permutation symmetries, and contracted with the total multipoles.
This procedure reproduces \crefrange{eq:Q00}{eq:Q44} and generalizes to additional multipoles without another manual derivation.
The tensor obeys
\[
 A_{\ell\ell';L_1L_2}
 =A_{\ell'\ell;L_1L_2}
 =A_{\ell\ell';L_2L_1}.
\]
For example, Legendre orthogonality gives
\[
 A_{00;22}=\frac12\int_{-1}^{1}\dd\mu\,\Lpoly_2^2(\mu)=\frac15,
\]
which is the coefficient of $S_2^2$ in \cref{eq:Q00}.

\paragraph{A worked shell.}
Take $b_1=2$, $f=0.8$, $\Plin=1000\,(h^{-1}\mathrm{Mpc})^3$, $\bar n_g=3\times10^{-4}\,(h\,\mathrm{Mpc}^{-1})^3$, $V=5\,(h^{-1}\mathrm{Gpc})^3$, $k=0.10\,h\,\mathrm{Mpc}^{-1}$, and $\Delta k=0.01\,h\,\mathrm{Mpc}^{-1}$.
Substituting into \crefrange{eq:P0kaiser}{eq:P4kaiser} gives
\begin{equation}
 (P_0,P_2,P_4)\simeq(5195,2499,146)\,(h^{-1}\mathrm{Mpc})^3,
 \qquad P_N=3333\,(h^{-1}\mathrm{Mpc})^3.
\end{equation}
For $N_{\rm full}\simeq2.53\times10^4$, contracting \cref{eq:Qgenerator} gives
\begin{equation}
 [\sigma(P_0),\sigma(P_2),\sigma(P_4)]
 \simeq(76,187,248)\,(h^{-1}\mathrm{Mpc})^3,
\end{equation}
and the correlation matrix
\begin{equation}
 R\simeq
 \begin{pmatrix}
 1&0.247&0.025\\
 0.247&1&0.209\\
 0.025&0.209&1
 \end{pmatrix}.
 \label{eq:numericalcorr}
\end{equation}
Although $P_4$ has the smallest mean, it has the largest absolute uncertainty.
The magnitude of a mean multipole therefore does not determine the magnitude of its fluctuations.

\section{Growing-mode spherical collapse and its perturbative limit}
\label{app:spherical-collapse}

For a bound patch, $E<0$.
Define $A\equiv-GM/(2E)$ and set $R=A(1-\cos\eta)$.
Substituting this parametrization into the energy equation gives
\begin{equation}
 \dot R^2=\frac{2GM}{R}-\frac{GM}{A},
 \qquad
 \frac{\dd t}{\dd\eta}
 =\frac{\dd R/\dd\eta}{\dot R}
 =\sqrt{\frac{A^3}{GM}}(1-\cos\eta)
 \equiv B(1-\cos\eta),
 \qquad A^3=GMB^2.
 \label{eq:cycloid-quadrature}
\end{equation}
The pure growing mode shares the background Big Bang time, so $t=0$ at $\eta=0$.
Integrating \cref{eq:cycloid-quadrature} then recovers the cycloid of \cref{eq:cycloid}.
Its early-time series is
\begin{align}
 1-\cos\eta&=\frac{\eta^2}{2}-\frac{\eta^4}{24}
 +\frac{\eta^6}{720}+\cdots,\nonumber\\
 \eta-\sin\eta&=\frac{\eta^3}{6}-\frac{\eta^5}{120}
 +\frac{\eta^7}{5040}+\cdots.
 \label{eq:cycloid-series}
\end{align}
Substituting these series into \cref{eq:deltanlsphere} first gives $\delta_{\rm NL}=3\eta^2/20+\Order(\eta^4)$.
In EdS the linear growing mode obeys $\delta_{\rm L}\propto t^{2/3}$.
Matching its leading term to $\delta_{\rm NL}$ therefore fixes
\[
 \delta_{\rm L}(\eta)
 =\frac{3}{20}\left[6(\eta-\sin\eta)\right]^{2/3}
 =\frac35\left[\frac34(\eta-\sin\eta)\right]^{2/3},
\]
which is \cref{eq:deltalsphere}.
Expanding both densities one order further gives
\begin{align}
 \delta_{\rm NL}(\eta)
 &=\frac{3}{20}\eta^2+\frac{37}{2800}\eta^4+\Order(\eta^6),\nonumber\\
 \delta_{\rm L}(\eta)
 &=\frac{3}{20}\eta^2\left(1-\frac{\eta^2}{30}
 +\frac{13\eta^4}{25200}+\cdots\right).
 \label{eq:spherical-series}
\end{align}
The common Big Bang time is equivalent to the pure growing-mode velocity fixed by \cref{eq:growing-mode-initialization};
a shifted Big Bang time would add a decaying mode and change the finite-time energy.
At formal collapse the nonlinear expression diverges, while the linearly extrapolated density remains finite:
\[
 \delta_c=\delta_{\rm L}(2\pi)
 =\frac{3}{20}(12\pi)^{2/3}\simeq1.686.
\]
Finally, eliminating $\eta^2$ between the two series gives
\begin{equation}
 \delta_{\rm NL}=\delta_{\rm L}+\frac{17}{21}\delta_{\rm L}^2+\cdots.
 \label{eq:sphereperturbative}
\end{equation}
The coefficient $17/21$ is the angle-averaged second-order density response in EdS and provides a direct consistency relation between the spherical model and perturbation theory.
Spherical collapse captures the isotropic part of nonlinear growth, whereas tidal operators and the Zel'dovich eigenvalues describe anisotropy.

\section{FFTLog evaluation and line-of-sight master integrals}
\label{app:master-integrals}

\subsection{FFTLog as a power-law basis}

On a logarithmic interval of length $L$, define $y=\ln(k/k_0)$.
Removing a real tilt $\nu_{\rm FFT}$ and Fourier expanding the approximately periodic remainder gives
\begin{equation}
 \Plin(k)=\sum_m c_m\left(\frac{k}{k_0}\right)^{\nu_{\rm FFT}+\ii\eta_m}
 =\sum_m c_m\left(\frac{k}{k_0}\right)^{-2\alpha_m},
 \qquad
 \alpha_m\equiv-\frac12(\nu_{\rm FFT}+\ii\eta_m).
 \label{eq:fftlogdecomp}
\end{equation}
The Fourier frequency is $\eta_m=2\pi m/L$.
An ordinary discrete FFT on a uniform $y$ grid determines the coefficients $c_m$.
They give a finite basis representation of the power spectrum on the transform interval.

\subsection{Scalar master integral and loop contractions}

In this basis every scalar loop monomial reduces to a single master integral over the two power-law factors produced by the expansion:
\begin{equation}
 I(\alpha,\beta;k)
 \equiv\intq\frac{1}{q^{2\alpha}|\vk-\vq|^{2\beta}}
 =k^{3-2(\alpha+\beta)}
 \frac{
 \Gamma(\tfrac32-\alpha)
 \Gamma(\tfrac32-\beta)
 \Gamma(\alpha+\beta-\tfrac32)}
 {8\pi^{3/2}\Gamma(\alpha)\Gamma(\beta)
 \Gamma(3-\alpha-\beta)}.
 \label{eq:scalarmaster}
\end{equation}
Dimensional analysis fixes the power of $k$, while a Feynman or Schwinger parameterization fixes the coefficient~\cite{FFTLog,FASTPT}.
Absolute convergence before analytic continuation requires
\begin{equation}
 \Re\alpha<\frac32,
 \qquad
 \Re\beta<\frac32,
 \qquad
 \Re(\alpha+\beta)>\frac32.
 \label{eq:scalarmaster-strip}
\end{equation}
The first two conditions control the local singularities at $\vq=0$ and $\vq=\vk$;
the third is the ultraviolet boundary as $q\to\infty$.
We then use analytic continuation for exponents outside this strip.
The individual gamma functions develop poles at special exponents even where the physical sum stays finite, and an implementation must treat these cancellations and the forbidden tilt values explicitly.

We derive the scalar master in the convergence strip of \cref{eq:scalarmaster-strip}.
Feynman parameterization followed by the shift $\vp=\vq-(1-x)\vk$ gives
\begin{equation}
 \frac{1}{A^\alpha B^\beta}
 =\frac{\Gamma(\alpha+\beta)}{\Gamma(\alpha)\Gamma(\beta)}
 \int_0^1\!\dd x\,
 \frac{x^{\alpha-1}(1-x)^{\beta-1}}
 {[xA+(1-x)B]^{\alpha+\beta}},
 \qquad
 \int\!\frac{\dd^3p}{(2\pi)^3}\frac{1}{(p^2+\Delta)^s}
 =\frac{\Gamma(s-\tfrac32)}{8\pi^{3/2}\Gamma(s)}
 \Delta^{3/2-s}.
 \label{eq:master-feynman}
\end{equation}
Taking $A=q^2$, $B=|\vk-\vq|^2$, and $\Delta=x(1-x)k^2$, the remaining $x$ integral is $B(\tfrac32-\beta,\tfrac32-\alpha)$.
Combining the beta and gamma functions reproduces \cref{eq:scalarmaster};
analytic continuation is applied only after this convergent derivation.

\begin{derivationbox}{FFTLog reduction of a kernel monomial}
After inserting two FFTLog modes, a scalar kernel term with numerator $q^{2u}|\vk-\vq|^{2v}$ becomes
\begin{equation}
 q^{2u}|\vk-\vq|^{2v}
 \frac{1}{q^{2\alpha_m}|\vk-\vq|^{2\alpha_n}}
 \longrightarrow I(\alpha_m-u,\alpha_n-v;k).
 \label{eq:fftlog-reduction}
\end{equation}
All dot products can be reduced to these shifts using
\[
 2\vk\cdot\vq=k^2+q^2-|\vk-\vq|^2.
\]
For $P_{22}$ both internal spectra are expanded, so the result is a double sum $\sum_{mn}c_mc_nM_{22}^{mn}$.
For $P_{13}$ one $\Plin$ factor is external and only the internal $\Plin(q)$ is expanded, so the result is $\Plin(k)\sum_m c_mM_{13}^m$, a vector contraction.
The double sum for $P_{22}$ and vector contraction for $P_{13}$ follow from the loop topology, not from the numerical method.
\end{derivationbox}

We expand $F_2$ and $F_3$ into scalar products and reduce every loop monomial with \cref{eq:fftlog-reduction}.
The finite kernel algebra reduces to analytic kernels in exponent space~\cite{FFTLog,FASTPT}.
These kernels do not depend on cosmology, so they can be tabulated once;
only the $c_m$ change with $\Plin$.
For the galaxy power spectrum, the same organization applies to every gravity, bias, velocity and redshift-space sector.
The number of matrices increases, but the method is unchanged.
For the IR-resummed spectrum of \cref{eq:IRassembled}, FFTLog evaluates the required full and no-wiggle loop inputs separately.
FFTLog does not define the physical split $\Plin=P_{\rm nw}+P_{\rm w}$;
that choice is made before either input is transformed.

\subsection{Line-of-sight tensor reduction}

Redshift-space kernels contain factors of $\nhat\cdot\vq$, whose angular form is restricted by rotational invariance.
Define the tensor family
\begin{equation}
 \mathcal I_N(\alpha,\beta;k,\mu)
 \equiv\intq\frac{(\nhat\cdot\vq)^N}
 {q^{2\alpha}|\vk-\vq|^{2\beta}}.
 \label{eq:LOSmasterdef}
\end{equation}
Rotational invariance and parity then force it to be a polynomial of degree $N$ in $\mu=\khat\cdot\nhat$ carrying the parity of $N$, and it takes the form
\begin{equation}
 \mathcal I_N
 =k^{3-2(\alpha+\beta)+N}
 \sum_{r=0}^{\lfloor N/2\rfloor}\mu^{N-2r}C_{N,r}(\alpha,\beta).
 \label{eq:LOSmasterpoly}
\end{equation}
To see why these are the only powers, first leave the line of sight uncontracted:
\begin{equation}
 T_{i_1\cdots i_N}
 =k^{3-2(\alpha+\beta)+N}
 \sum_{r=0}^{\lfloor N/2\rfloor}A_{N,r}\,
 \mathcal S\!\left[
 \delta_{i_1i_2}\cdots\delta_{i_{2r-1}i_{2r}}
 \hat k_{i_{2r+1}}\cdots\hat k_{i_N}\right],
 \label{eq:tensor-decomposition}
\end{equation}
where $\mathcal S$ symmetrizes the indices.
Rotational invariance permits only Kronecker delta pairs and factors of $\khat$.
Contracting with $n_{i_1}\cdots n_{i_N}$ turns every delta pair into $n_in_j\delta_{ij}=1$ and each remaining $\khat$ into $n_i\hat k_i=\mu$.
Thus a term with $r$ pairs produces
\[
 \underbrace{1\cdots1}_{r\ {\rm pairs}}
 \underbrace{\mu\cdots\mu}_{N-2r\ {\rm factors}}=\mu^{N-2r},
\]
with the symmetrization multiplicity absorbed into $C_{N,r}$.
Ref.~\cite{ClassPT} builds these redshift-space master integrals in its Appendix~A by writing each $C_{N,r}$ as a combination of the scalar integral of \cref{eq:scalarmaster} at shifted exponents.
A source generating function supplies the same coefficients in closed form,
\begin{equation}
 \begin{split}
 C_{N,r}(\alpha,\beta)={}&
 \frac{N!}{r!(N-2r)!4^r}\\
 &\times
 \frac{
 \Gamma(\alpha+\beta-\tfrac32-r)
 \Gamma(\tfrac32-\beta+r)
 \Gamma(N-r+\tfrac32-\alpha)}
 {8\pi^{3/2}\Gamma(\alpha)\Gamma(\beta)
 \Gamma(N+3-\alpha-\beta)}.
 \end{split}
 \label{eq:LOScoeff}
\end{equation}
The first members of the family are
\begin{align}
 \mathcal I_1&=k^{4-2(\alpha+\beta)}\mu A_1,
 &\mathcal I_2&=k^{5-2(\alpha+\beta)}(A_2+\mu^2B_2),\\
 \mathcal I_3&=k^{6-2(\alpha+\beta)}(\mu A_3+\mu^3B_3),
 &\mathcal I_4&=k^{7-2(\alpha+\beta)}(A_4+\mu^2B_4+\mu^4D_4).
\end{align}
In terms of \cref{eq:LOScoeff}, $A_1=C_{1,0}$;
$A_2=C_{2,1}$ and $B_2=C_{2,0}$;
$A_3=C_{3,1}$ and $B_3=C_{3,0}$;
and $A_4=C_{4,2}$, $B_4=C_{4,1}$ and $D_4=C_{4,0}$.
Within each rank, the letters ascend with the power of $\mu$ while $r$ descends.
We write $D_4$ in place of $C_4$ to avoid a collision with the two-index family.
These coefficient families encode the tensor structures allowed by the symmetries.
Matter in real space, matter in redshift space and biased tracers then correspond to different truncations of one hierarchy of master integrals.

For the tensor numerator, introduce a one-dimensional source along the line of sight and use Schwinger parameters:
\begin{equation}
 \begin{split}
 \mathcal G(z)&\equiv\intq
 \frac{e^{z\nhat\cdot\vq}}{q^{2\alpha}|\vk-\vq|^{2\beta}}\\
 &=\frac{1}{8\pi^{3/2}\Gamma(\alpha)\Gamma(\beta)}
 \int_0^\infty\!\dd s\,\dd t\,
 \frac{s^{\alpha-1}t^{\beta-1}}{(s+t)^{3/2}}
 \exp\!\left[-\frac{st}{s+t}k^2
 +\frac{t}{s+t}zk\mu+\frac{z^2}{4(s+t)}\right],\\
 \mathcal I_N&=\left.\frac{\dd^N\mathcal G}{\dd z^N}\right|_{z=0}.
 \end{split}
 \label{eq:los-generating-source}
\end{equation}
The source derivative separates the contractions associated with the $r$ Kronecker-delta pairs from the remaining line-of-sight factors:
\[
 \left.\frac{\dd^N}{\dd z^N}e^{az+bz^2}\right|_{z=0}
 =N!\sum_{r=0}^{\lfloor N/2\rfloor}
 \frac{a^{N-2r}b^r}{(N-2r)!r!}.
\]
Set $s=S(1-y)$ and $t=Sy$.
For the term with $r$ pairs, the two remaining integrals are
\begin{align*}
 \int_0^\infty\!\dd S\,S^{\alpha+\beta-5/2-r}
 e^{-Sy(1-y)k^2}
 &=\frac{\Gamma(\alpha+\beta-\tfrac32-r)}
 {[y(1-y)k^2]^{\alpha+\beta-3/2-r}},\\
 \int_0^1\!\dd y\,
 y^{N-r+1/2-\alpha}(1-y)^{r+1/2-\beta}
 &=B\!\left(N-r+\tfrac32-\alpha,\tfrac32-\beta+r\right).
\end{align*}
Together with the derivative multiplicity, these factors give exactly $C_{N,r}$ in \cref{eq:LOScoeff} and the common power $k^{3-2(\alpha+\beta)+N}$.

Two low-rank checks verify the derivation.
For $N=0$,
\begin{equation}
 k^{3-2(\alpha+\beta)}C_{0,0}(\alpha,\beta)
 =I(\alpha,\beta;k),
 \label{eq:los-n0-check}
\end{equation}
because $C_{0,0}$ is exactly the gamma-function coefficient of \cref{eq:scalarmaster}.
For $N=2$, the tensor has the form $T_{ij}=k^{5-2(\alpha+\beta)}(C_{2,1}\delta_{ij}+C_{2,0}\hat k_i\hat k_j)$.
Tracing it either contracts $q_iq_i=q^2$ in the integral or contracts the tensor basis, so
\begin{equation}
 k^{5-2(\alpha+\beta)}\left(3C_{2,1}+C_{2,0}\right)
 =I(\alpha-1,\beta;k).
 \label{eq:los-n2-trace}
\end{equation}
The equality tests both the normalization and the pair multiplicity.

\paragraph{Minimal numerical checks.}
Reconstructing $\Plin$ from its FFTLog coefficients tests the basis representation.
Varying $\nu_{\rm FFT}$, the transform range, padding and grid size tests numerical convergence.
Selected $P_{22}$ and $P_{13}$ contributions should agree with independent direct quadrature, and the final multipoles should match an independent implementation~\cite{ClassPT,velocileptors,Beutler}.
These checks diagnose numerical error;
EFT power counting or an omitted-order estimate diagnoses perturbative truncation.

\section{Answer sketches for the exercises}
\label{app:answers}

These sketches identify the intended method but omit enough intermediate algebra to preserve the exercises as independent calculations.

\Needspace{7\baselineskip}
\paragraph{\hyperref[exercises:lecture1]{Lecture 1}.}
\begin{enumerate}
\item With $a\propto\tau^2$, $\cH=2/\tau$, and $\Omega_m=1$, \cref{eq:growth} becomes $\delta''+(2/\tau)\delta'-(6/\tau^2)\delta=0$.
Substituting $\delta\propto\tau^n$ gives $(n-2)(n+3)=0$.
Thus $D_+\propto\tau^2\propto a$ and $D_-\propto\tau^{-3}\propto a^{-3/2}$.
\item Insert both inverse transforms:
\begin{align*}
 \sigma_R^2
 &=\int\!\frac{\dd^3k}{(2\pi)^3}
 \int\!\frac{\dd^3k'}{(2\pi)^3}
 e^{\ii(\vk+\vk')\cdot\vx}W(kR)W(k'R)
 \langle\delta(\vk)\delta(\vk')\rangle \\
 &=\int\!\frac{\dd^3k}{(2\pi)^3}
 W(kR)W(-\vk R)P(k).
\end{align*}
Here \cref{eq:powerdef} sets $\vk'=-\vk$.
Reality of the real-space window gives $W(-\vk R)=W^*(\vk R)$, while spherical symmetry makes $W(\vk R)=W(kR)$.
Therefore
\[
 \sigma_R^2=\int\frac{\dd k\,k^2}{2\pi^2}P(k)|W(kR)|^2
 =\int\dd\ln k\,\Delta^2(k)|W(kR)|^2,
\]
as required by \cref{eq:smoothing-variance}.
\item In units with $c=1$, $k^2/H_0^2$ is dimensionless.
The two minus signs in \cref{eq:curvature-to-density} cancel, so the correct kernel is positive;
with $D_+(a)=D_+(1)\widehat D_+(a)$,
\[
 \frac{\mathcal M_{\rm code}}{\mathcal M}=-\frac{1}{D_+(1)}.
\]
The sign cancels from the auto-power but reverses $\delta_m/\mathcal R$.
The missing constant $D_+(1)$ changes the auto-power amplitude as well as the field normalization.
\end{enumerate}

\paragraph{\hyperref[exercises:lecture2]{Lecture 2}.}
\begin{enumerate}
\item At turnaround, $\eta=\pi$ gives $\delta_{\rm L}=\tfrac35(3\pi/4)^{2/3}\simeq1.06$, and at collapse $\eta=2\pi$ gives $\delta_c=\tfrac35(3\pi/2)^{2/3}=\tfrac{3}{20}(12\pi)^{2/3}\simeq1.686$.
The nonlinear density diverges as the cycloid reaches zero radius;
the linearly extrapolated $\delta_{\rm L}$ remains finite.
\item Expanding gives $\delta_{\rm ZA}=D_+\operatorname{tr}d+\Order(D_+^2)$.
\cref{eq:za-deformation} gives $\nabla_q\cdot\vPsi^{(1)}=-D_+\operatorname{tr}d$.
By \cref{eq:psilinear}, $D_+\operatorname{tr}d=\delta^{(1)}$.
Thus $\delta_{\rm ZA}=\delta^{(1)}+\Order(D_+^2)$ at first order.
\item Differentiating \cref{eq:threshold-tail} at $\delta_L=0$ gives
\[
 \begin{aligned}
 \left.\frac{\partial F}{\partial\delta_L}\right|_0
 &=\frac{e^{-\nu_c^2/2}}{\sqrt{2\pi}\,\sigma},\\
 b_1^{\rm L}
 &=\bar F^{-1}\left.\frac{\partial F}{\partial\delta_L}\right|_0.
 \end{aligned}
\]
The supplied large-tail asymptotic then gives
\[
 b_1^{\rm L}\simeq
 \frac{e^{-\nu_c^2/2}}{\sqrt{2\pi}\,\sigma}
 \frac{\sqrt{2\pi}\,\nu_c}{e^{-\nu_c^2/2}}
 =\frac{\nu_c}{\sigma}
 =\frac{\nu_c^2}{\delta_c}.
\]
Rare high-$\nu_c$ objects therefore have a large fractional abundance response to a long-wavelength density perturbation.
\item From \cref{eq:growing-mode-initialization}, the kinetic and potential terms are $K_{H,i}(1-2\delta_i/3)$ and $K_{H,i}(1+\delta_i)$, so $E_i=-(5/3)K_{H,i}\delta_i$.
Thus $A\propto\delta_i^{-1}$ and $A^3=GMB^2$ gives $B\propto\delta_i^{-3/2}$ and $t_{\rm coll}=2\pi B$.
Since linear EdS growth is proportional to $t^{2/3}$, $\delta_i(t_{\rm coll}/t_i)^{2/3}\propto\delta_i\delta_i^{-1}$, leaving a universal $\delta_c$.
\end{enumerate}

\paragraph{\hyperref[exercises:lecture3]{Lecture 3}.}
\begin{enumerate}
\item Gram--Schmidt first gives
\[
 u_2=\mu^2-\frac13,
 \qquad
 u_4=\mu^4-\frac15\Lpoly_0-\frac47\Lpoly_2.
\]
Imposing $\Lpoly_\ell(1)=1$ fixes the standard normalization,
\[
 \Lpoly_0=1,
 \qquad
 \Lpoly_2=\frac32u_2=\frac12(3\mu^2-1),
 \qquad
 \Lpoly_4=\frac{35}{8}u_4
 =\frac18(35\mu^4-30\mu^2+3).
\]
Substituting these polynomials and \cref{eq:kaiserexpanded} directly into the projection integral gives
\[
 P_0=\left(b_1^2+\frac23b_1f+\frac15f^2\right)\Plin,
 \quad
 P_2=\left(\frac43b_1f+\frac47f^2\right)\Plin,
 \quad
 P_4=\frac{8}{35}f^2\Plin,
\]
recovering \crefrange{eq:P0kaiser}{eq:P4kaiser}.
\item With $S_2=S_4=0$, \cref{eq:Qdef} and Legendre orthogonality give
\[
 Q_{\ell\ell'}
 =\frac{(2\ell+1)(2\ell'+1)}{2}S_0^2
 \frac{2\delta^K_{\ell\ell'}}{2\ell+1}
 =(2\ell+1)S_0^2\delta^K_{\ell\ell'},
 \qquad S_0=P_0+P_N.
\]
Thus every diagonal estimator has sampling variance, even when its mean multipole vanishes, while orthogonality makes all off-diagonal blocks zero for an isotropic total power.
\item For a zero-mean Gaussian field in a narrow isotropic shell, take $P_{\rm tot}$ constant across the shell and let $\sigma^2=P_{\rm tot}/2$.
For independent Gaussian $X$ and $Y$, $\langle|\delta|^2\rangle=2\sigma^2=P_{\rm tot}$ and
\[
 \langle|\delta|^4\rangle
 =\langle(X^2+Y^2)^2\rangle
 =8\sigma^4=2P_{\rm tot}^2.
\]
Hence $\operatorname{Var}|\delta|^2=P_{\rm tot}^2$.
Averaging independent modes gives
\[
 \operatorname{Var}\widehat P_0
 =\frac{P_{\rm tot}^2}{N_i^{\rm ind}}
 =\frac{2P_{\rm tot}^2}{N_i^{\rm full}}.
\]
Subtracting known $P_N$ is a constant shift and changes only the mean.
If shot noise is estimated rather than known, its uncertainty adds covariance.
\end{enumerate}

\paragraph{\hyperref[exercises:extension]{Extension}.}
\begin{enumerate}
\item Expanding the exponential in \cref{eq:rsdexact}, the product of the first-order density with the linear mapping term gives
\[
 \frac{fk\mu}{2}\left(\frac{\mu_1}{k_1}+\frac{\mu_2}{k_2}\right).
\]
The quadratic exponential term is $f^2(k\mu)^2\mu_1\mu_2/(2k_1k_2)$, where $k\mu=k_1\mu_1+k_2\mu_2$.
Combining the two terms gives
\[
 \frac{fk\mu}{2}\left[
 \frac{\mu_1}{k_1}(1+f\mu_2^2)
 +\frac{\mu_2}{k_2}(1+f\mu_1^2)\right].
\]
The two terms exchange under $\vk_1\leftrightarrow\vk_2$, so the bracket is symmetric.
\item Mass and momentum conservation force $P_{\rm stoch}^m=\Order(k^4)$, so the stochastic matter power has no white-noise term.
Discrete tracer number is not constrained by local matter rearrangements and may retain constant stochastic power.
For a localized rearrangement $\Delta\rho(\vx)$, define
\[
 \delta_{\rm stoch}^m(\vk)
 =\frac{1}{\bar\rho_m}\int\dd^3x\,e^{-\ii\vk\cdot\vx}\Delta\rho(\vx).
\]
Expanding at low $k$ gives
\[
 \delta_{\rm stoch}^m(\vk)
 =\frac{1}{\bar\rho_m}\left[
 \int\Delta\rho
 -\ii k_i\int x_i\Delta\rho
 -\frac12k_ik_j\int x_ix_j\Delta\rho+\cdots\right].
\]
Mass conservation removes the monopole, and momentum conservation removes the dipole.
Therefore $\delta_{\rm stoch}^m=\Order(k^2)$ and $P_{\rm stoch}^m=\Order(k^4)$.
\end{enumerate}

\bibliography{refs}

\end{document}